\documentclass[%
 aip,
 amsmath,amssymb,
]{revtex4-1}

\usepackage{graphicx}
\usepackage{dcolumn}
\usepackage{bm}
\usepackage{siunitx}
\usepackage{enumitem}
\usepackage{xcolor}

\begin{document}


\title{Effect of spatial distribution of mesoscale heterogeneities\\on the shock-to-detonation transition in liquid nitromethane}

\author{XiaoCheng Mi}
 \email{xcm20@cam.ac.uk; xiaocheng.mi@mail.mcgill.ca}
\author{Louisa Michael}
\author{Nikolaos Nikiforakis}%
\affiliation{Cavendish Laboratory, Department of Physics, University of Cambridge, Cambridge, United Kingdom CB3 0HE}

\author{Andrew J. Higgins}
\affiliation{Department of Mechanical Engineering, McGill University, Montreal, Quebec, Canada H3A 0C3}%

\date{\today}

\begin{abstract}
The sensitizing effect of cavities in the form of microbubbles on the shock initiation of a homogeneous liquid explosive is studied computationally. While the presence of voids in an explosive has long been known to induce so-called hot spots that greatly accelerate the global reaction rate, the ability to computationally resolve the details of the interaction of the shock front with heterogeneities existing on the scale of the detonation reaction zone has only recently become feasible. In this study, the influence of the spatial distribution of air-filled cavities has been examined, enabled by the use of graphic processing unit (GPU) accelerated computations that can resolve shock initiation and detonation propagation through an explosive while fully resolving features at the mesoscale. Different spatial distributions of cavities are examined in two-dimensional simulations, including regular arrays of cavities, slightly perturbed arrays, random arrays (with varying minimum spacing being imposed on the cavities), and randomly distributed clusters of cavities. Statistical ensembles of simulations are performed for the cases with randomly positioned cavities. The presence of the cavities is able to reduce the time required to initiate detonation—for a given input shock strength---by greater than $50\%$, in agreement with previous experimental results. Randomly distributing the cavities results in a $15$-$20\%$ decrease in detonation initiation time in comparison to a regular array of cavities. Clustering the cavities---as would occur in the case of agglomeration---results in an additional $10\%$ decrease in detonation initiation time in comparison to random arrays. The effect of clustering is shown not to be a result of the clusters forming an effectively larger cavity, but rather due to interactions between clusters upon shock loading occurring on the microscale.  The implications of these results for modelling and experiments of microbubble-sensitized explosives is discussed. 
\end{abstract}


\maketitle

\section{\label{Intro}Introduction}
The shock-to-detonation transition (SDT) in a condensed-phase explosive is accelerated by the presence of mesoscale heterogeneities. The interaction between an incident shock wave and heterogeneities creates localised high-temperature regions, i.e., ``hot spots'', from where reaction waves emanate and evolve into a detonation~\cite{BowdenYoffe1948hot}.  The formation and evolution of hot spots are linked to the statistical nature of the mesoscale heterogeneities, controlling the macroscopic SDT behaviour~\cite{Baer2002}. However, the detailed mechanisms underlying the collective effect of a large number of hot spots are yet to be fully understood.\\

In solid explosives, the intrinsic mesoscale heterogeneities are highly irregular in size, shape, and spatial distribution. The formation of hot spots in solid explosives is attributed to a variety of mechanisms, including viscoplastic heating, shear banding, intergranular friction, etc.~\cite{Field1992hot,Solovev2000}. In order to probe the fundamental mechanisms of hot-spot formation and evolution, liquid explosives with artificially added inert inclusions has been a preferred system by researchers to carry out better controlled investigations. Gelled nitromethane (NM) mixed with silica beads or glass micro-balloons (GMBs) have been used to experimentally examine the sensitising effect of mesoscale heterogeneities on an SDT process~\cite{Engelke1979,Engelke1983,Presles1995,Gois1996, Gois1996APS,Gois2002,Sheffield1989Report,Dattelbaum2009APS,Dattelbaum2010,Dattelbaum2010Role,Higgins2018APS}. In such systems with a highly viscous liquid explosive, the volume (or weight) fraction and size distribution of the heterogeneities are relatively controllable; the formation mechanism of a single hot spot, i.e., shock reflection upon a piece of inert inclusion, is relatively well understood (with the aid of computational simulations~\cite{Ball2000,Swantek2010JFM,Menikoff2011SWJ,Hawker2012,Lauer2012,Ozlem2012,Kapila2015,Betney2015,Apazidis2016,Michael2018_I,Michael2018_II,Michael2019Book}). It is however difficult to experimentally reveal how the intrinsic structure of mesoscale heterogeneities (on the order of \SI{1}-\SI{100}{\micro\metre}) influences the collective evolution of hot spots, and thus, determine the macroscopic SDT behaviour over a distance of millimetres or centimetres. Computational modelling has otherwise been used to gain more insights into the mesoscale dynamics of hot spots.\\

With advanced computing technologies, simulations explicitly resolving a statistically significant number of mesoscale heterogeneities, i.e., meso-resolved simulations, have recently become feasible and been used to model shock-initiation phenomena in solid explosives~\cite{Rai2015JAP,Kim2018JMPS,Yarrington2018JAP,JacksonJost2018CTM,Nassar2019,Roy2019,KimYoh2019CNF,Wei2019,Zhang2019SWJ,Miller2019PEP,JacksonZhangShort2019PEP}. By performing meso-resolved simulations of liquid NM with air-filled cavities without invoking any phenomenological reaction rate model, Mi~\textit{et al}.~\cite{Mi2019JAP} captured the characteristic SDT behaviour in a heterogeneous explosive, demonstrating the sensitising effect of the heterogeneities, i.e., a significant reduction in the detonation overtake time compared to that for the case of neat NM (without any cavities). Another finding of~\cite{Mi2019JAP} is that a random distribution of cavities results in a more pronounced sensitising effect than a regular array of cavities does. This finding suggests that, for a fixed overall porosity, the resulting SDT behaviour differs as the nature of the spatial distribution of mesoscale heterogeneities is varied. Using this relatively simple system of liquid-phase heterogeneous explosive, the current work is an attempt to further understand the relation between the statistical nature of heterogeneity distribution and the collective hot-spot effect on an SDT process. Numerical experiments are carried out to examine the sensitising effect resulting from various idealized spatial distributions of heterogeneities.\\
\begin{figure*}
\centerline{\includegraphics[width=0.8\textwidth]{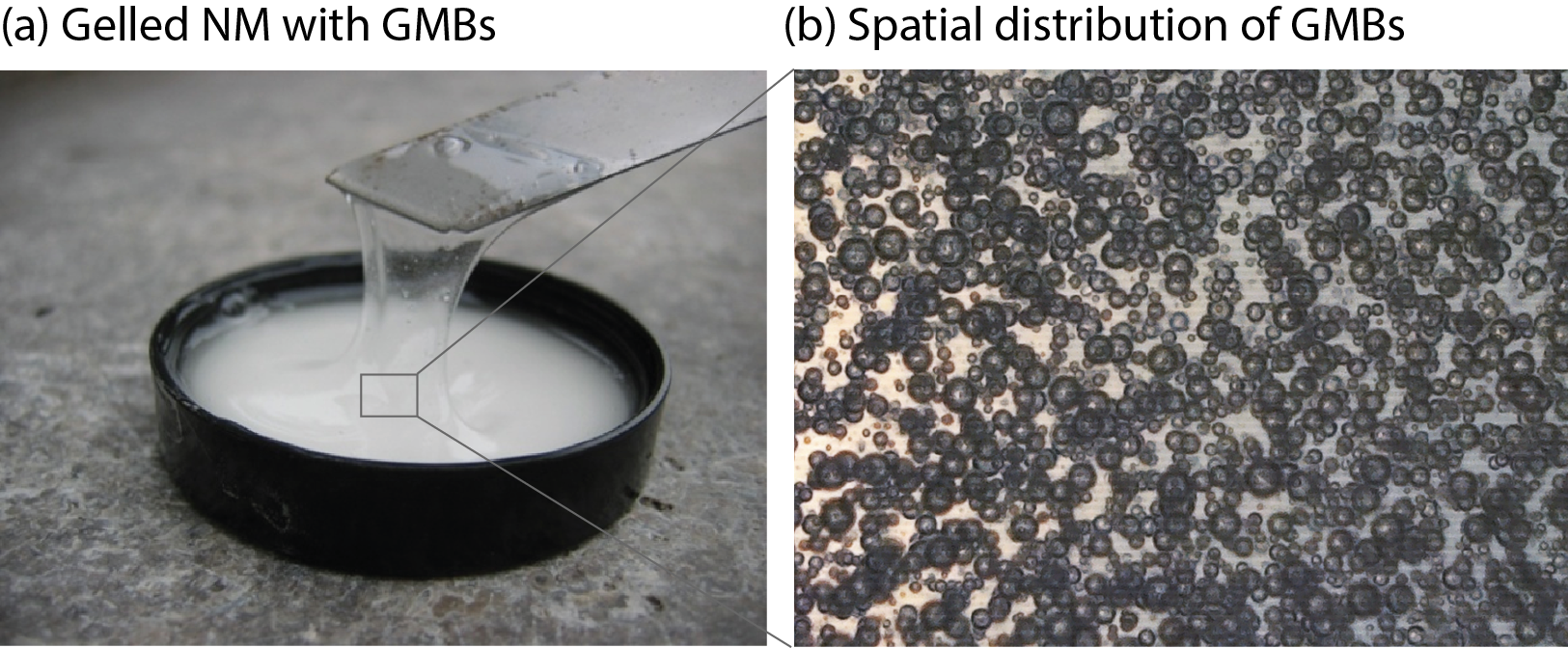}}
		\caption{(a) A sample mixture of PMMA-gelled nitromethane and glass micro-balloons (GMBs) and (b) the spatial distribution of GMBs at mesoscale.}
	\label{NM_sample}
\end{figure*}

As revealed in a number of experimental studies~\cite{Dattelbaum2010,Higgins2018APS}, the spatial distribution of mesoscale heterogeneities in liquid explosives is highly non-uniform. Due to a high viscosity of a gelled liquid explosive matrix, flows over mesoscale (micron-sized) inert inclusions are of very low Reynolds numbers. Experimental observation and theoretical analysis demonstrate that, in a suspension of more than three closely located particles at low Reynolds numbers (of the order of $0.01$), clusters naturally form under the unbalanced hydrodynamic forces exerted on each particle~\cite{Happel1983,Crowley1971,Crowley1976}. Evidence of clusters of glass micro-balloons (GMBs), as shown in Fig.~\ref{NM_sample}, can be found in a sample mixture with PMMA-gelled (Polymethyl methacrylate-gelled) NM used in experimental studies~\cite{Higgins2018APS}. In order to tailor the blasting performance of commercial explosives used for various mining applications, gas-filled voids are introduced into liquid (or emulsion) explosives with a non-uniform distribution, i.e., with regions wherein voids are highly concentrated and regions of a diluted concentration of voids~\cite{Johann2018Patent}. In the present study, the SDT behaviour resulting from an idealized representation of clusters of air-filled cavities is compared to those resulting from the cases with a regular array and a random distribution of cavities. Further, a spectrum of spatial distributions from a regular array to a random distribution (which are created via introducing small perturbations to a regular array or imposing a minimum spacing between neighbouring cavities to a random distribution) is examined to probe how the statistical nature of the mesoscale morphology influences the SDT behaviour. Statistical analysis, i.e., calculating the probability density function (PDF) of the energy release rate over space and time, is performed to illustrate how the hot-spot sensitization mechanism differs in various scenarios of cavity distribution.\\

This paper is organized as follows. In Sect.~\ref{sec2}, the governing equations and various spatial distributions of cavities considered in the simulations are described. Section~\ref{sec3} presents the herein used numerical methodology in detail. Simulation results and analysis are reported in Sect.~\ref{sec4}. The findings based on the results are discussed in Sect.~\ref{sec5} and summarized in Sect.~\ref{sec6}.

\section{\label{sec2}Problem description}

\subsection{Governing equations}
\label{sec2_1}
The dynamic system at hand is simulated in an Eulerian framework using a form of the MiNi16 mathematical formulation~\cite{Michael2016JCP} for liquid NM and air as the two immiscible materials in the multiphase model. The use of this mathematical framework has previously been explored by Michael and Nikiforakis to simulate collapse of a single cavity in liquid NM~\cite{Michael2014DS,Michael2018_I,Michael2018_II,Michael2019Book}. The simulations reported in this paper are performed in two-dimensions. The governing equations are formulated as follows,
\begin{equation}
\begin{split}
\frac{\partial z_1 \rho_1}{\partial t} + \nabla \cdot \left(z_1 \rho_1 \mathbf{u} \right) = 0 \\
\frac{\partial z_2 \rho_2}{\partial t} + \nabla \cdot \left(z_2 \rho_2 \mathbf{u} \right) = 0 \\
\frac{\partial}{\partial t}\left( \rho \mathbf{u} \right) + \nabla \cdot \left(\rho \mathbf{u} \otimes \mathbf{u} \right) + \nabla p = 0 \\
\frac{\partial}{\partial t}\left( \rho E \right) + \nabla \cdot \left[ \left(\rho E + p \right) \mathbf{u} \right] = -z_2 \rho_2 K Q\\
\frac{\partial z_1}{\partial t} + \mathbf{u} \cdot \nabla z_1 = 0 \\
\frac{\partial}{\partial t} \left(z_2 \rho_2 \lambda \right) + \nabla \cdot \left(z_2 \rho_2 \lambda \mathbf{u} \right) = z_2 \rho_2 K
\end{split}
\label{Eq1}
\end{equation}
The air within cavities is considered as phase~$1$ and the liquid NM is considered as phase~2, which are denoted as subscripts ``$1$'' and ``$2$'', respectively. The volume fractions of air and NM are represented by $z_1$ and $z_2$, respectively, where $z_1 + z_2 = 1$. The total density $\rho$ of the mixture can thus be calculated as $\rho = z_1 \rho_1 + z_2 \rho_2$. The total specific energy is defined as $E = \frac{1}{2}u^2 + e$, where $e$ is the specific internal energy of the mixture, i.e., $\rho e = z_1 \rho_1 e_1 + z_2 \rho_2 e_2$. The variable $\lambda$ represents the mass fraction of NM reactant, evolving from $1$ to $0$ as reactant is depleted through the chemical reaction. The time rate of change of $\lambda$ is denoted as $K$. The specific energy release of NM is represented by $Q$. At material interfaces, these two immiscible materials are considered to be in mechanical equilibrium (i.e., $p_1=p_2=p$, $\mathbf{u}_1=\mathbf{u}_2=\mathbf{u}$), but not necessarily in thermal equilibrium (i.e., $T_1$ and $T_2$ do not need to be equal). Since relatively high incident shock pressures ($> 7~\mathrm{GPa}$) are considered in this study, the time scale of the SDT process is expected to be on the order of~$\SI{}{\micro\second}$. Viscous and thermal diffusion is therefore neglected in this model. In an explosive mixture consisting of only liquid and gaseous materials, the effect of viscous heating is much less significant than that in solid heterogeneous explosives, which have an orders-of-magnitude greater viscosity~\cite{MenikoffSewell2002}. Thus, neglecting viscous heating and considering that the formation and growth of hot spots are dominated by hydrodynamic processes are reasonable assumptions for a mixture of liquid NM and air under detonation shock loading.\\

Air inside the cavities is governed by the ideal gas law, i.e., $e_1 = p/(\gamma_1 - 1)\rho_1$, where $\gamma_1$ is the ratio of specific heat capacities of air. The unreacted NM is described by the Cochran-Chan Equation of State (EoS)~\cite{Cochran1979,Saurel2009JCP}, which is expressed in a Mie-Gr\"{u}neisen form as,
\begin{equation}
p(\rho_2, e_2) = p_\mathrm{ref,2}(\rho_2) + \rho_2 \Gamma_{0,2} \left[e_2-e_\mathrm{ref,2}(\rho_2) \right]
\label{Eq2}
\end{equation}
where the reference pressure $p_\mathrm{ref,2}$ is given by
\begin{equation}
p_\mathrm{ref,2}(\rho_2) = \mathcal{A} \left(\frac{\rho_{0,2}}{\rho_2}\right)^{-\mathcal{C}} -\mathcal{B} \left(\frac{\rho_{0,2}}{\rho_2}\right)^{-\mathcal{D}}
\label{Eq3}
\end{equation}
the reference energy $e_\mathrm{ref,2}$ is given by
\begin{equation}
\begin{split}
e_\mathrm{ref,2}(\rho_2) = & \frac{-\mathcal{A}}{\rho_{0,2}(1-\mathcal{C})} \left[ \left(\frac{\rho_{0,2}}{\rho_2}\right)^{1-\mathcal{C}} -1 \right] \\
+ & \frac{\mathcal{B}}{\rho_{0,2}(1-\mathcal{D})} \left[ \left(\frac{\rho_{0,2}}{\rho_2}\right)^{1-\mathcal{D}} -1 \right]
\end{split}
\label{Eq4}
\end{equation}
and $\Gamma_{0,2}$ is the Gr\"{u}neisen coefficient corresponding to the initial state of NM at $\rho_2 = \rho_{0,2}$. This EoS has been broadly used in the literature to model unreacted NM~\cite{Saurel2009JCP,Shukla2010,Genetier2014,Michael2018_I,Michael2018_II,Michael2019Book}. For simplicity, the same EoS is used to approximately represent the reaction products of NM. Although the quantitative accuracy of this simplified model in comparison to experimental data might be compromised, a comparative study examining the qualitative effect of different mesoscale distributions on the SDT process is unlikely affected.\\

A general expression for calculating the temperature of each material based on the first-order Taylor expansion from the reference curve should be as follows
\begin{equation}
    T_i - T_{\mathrm{ref},i}(\rho_i) = \frac{p - p_{\mathrm{ref},i}(\rho_i)}{\rho_i \Gamma_{i}(\rho_i,T_i) c_{v,i}(\rho_i,T_i)}  \;\;\;\;\;\;\mathrm{for} \;\;i=1,2
    \label{Eq5}
\end{equation}
where the constant-volume specific heat capacity $c_{v,i}$ and  Gr\"{u}neisen coefficient $\Gamma_{i}$ are functions of both density and temperature~\cite{Winey2000}.  The Cochran-Chan EoS used for liquid NM is based on the reference curves of isotherms \cite{Cochran1979} imposing a temperature $T_2 = \SI{298}{K}$ at the initial density of NM. Thus, the Cochran-Chan isothermal reference curves are of a reference temperature $T_{\mathrm{ref},2} = \SI{0}{K}$. With a constant Gr\"{u}neisen parameter $\Gamma_2$ and specific heat capacity $c_{v,2}$, the expression for calculating the temperature of NM in this model is given as follows,
\begin{equation}
    T_2 = \frac{p - p_{\mathrm{ref},2}(\rho_2)}{\rho_2 \Gamma_{2} c_{v,2}} 
    \label{Eq6}
\end{equation}
The values of the EoS parameters for air and liquid NM are provided in Tables~\ref{Tab1} and \ref{Tab2}, respectively. A more detailed justification of the selected EoS for liquid NM can be found in~\cite{Mi2019JAP}. The reaction rate $K$ in Eq.~\ref{Eq1} is governed by single-step Arrhenius kinetics as follows,
\begin{equation}
K = \frac{\partial \lambda}{\partial t} = -\lambda C \mathrm{exp} (-T_\mathrm{a}/T_2)
\label{Eq7}
\end{equation}
where $T_\mathrm{a}$ is the activation temperature and $C$ is the pre-exponential factor. Note that, as $\lambda$ represents the mass fraction of NM reactant, the reaction rate $K$ is negative throughout the reaction. The values of $T_\mathrm{a}$, $C$, and $Q$ are summarized in Table~\ref{Tab3}. A detailed description of how these values were calibrated can be found in~\cite{Mi2019JAP}. Note that the complex chemical kinetics underlying NM decomposition at elevated pressures ($\sim~\SI{}{\giga\pascal}$) and temperatures cannot be adequately described by a single-step Arrhenius rate law.  Shaw \textit{et al.}~\cite{Shaw1979} found that the thermal explosion times of NM at pressures from $0.1$ to $\SI{5}{\giga\pascal}$ cannot be fitted to a function of temperature with a single value of activation energy. More recently, via VISAR (Velocity Interferometer System for Any Reflector) measurement of particle velocity, Bouyer~\textit{et al.}~\cite{Bouyer2009APS} revealed that a detonation wave in liquid NM consists of a zone of fast reaction followed by a zone of slow reaction. However, as opposed to the detailed chemical kinetics for NM reaction at atmospheric pressure, there are no well-determined rate constants or detailed pathways at elevated pressures that can be applied to SDT in liquid NM. Moreover, there is no valid EoS for any intermediate species of NM decomposition at elevated pressures. Given all of these uncertainties in incorporating a more detailed reaction model, to simulate NM detonation and SDT phenomena, a single-step Arrhenius reaction model with calibrated rate constants is commonly used.~\cite{Nunziato1983,Menikoff2011}\\

With the selected EoS and reaction rate model, as shown in~\cite{Mi2019JAP}, the detonation overtake times resulting from the one-dimensional simulations for neat NM quantitatively well agree with the experimental for input shock pressures less than $\SI{9.0}{\giga\pascal}$. Since a range of low input shock pressures (from $7.0$ to $\SI{8.2}{\giga\pascal}$) is considered in the present study to examine hot-spot-driven SDT processes, the captured reaction time in bulk NM is expected to be of a realistic scale. It is of importance to clarify that the focus of this study is \emph{not} targeted at an improved set of constitutive relations to achieve a quantitatively better match with the experimental data of a specific NM-based heterogeneous explosive. Rather, using this relatively simple EoS and reaction rate model for NM that were validated for shock-cavity interactions in previous studies~\cite{Michael2018_I,Michael2018_II,Mi2019JAP}, a series of systematic numerical experiments is devised in order to examine how the statistical nature of mesoscale heterogeneities affects the macroscopic SDT behavior.\\

\begin{table}
\begin{center}
\caption{Parameters for the equation of state of air}
\label{Tab1}
\begin{tabular}{| c | c |}
\hline
Parameter & Value (unit)\\
\hline
$\gamma_1$ & $1.4$ (-)\\
$c_{v,1}$ & $718$ ($\mathrm{J}$ $\mathrm{kg}^{-1}\mathrm{m}^{-3}$)\\
\hline
\end{tabular}
\end{center} 
\end{table}

\begin{table}
\begin{center}
\caption{Parameters for the equation of state of liquid NM}
\label{Tab2}
\begin{tabular}{| c | c |}
\hline
Parameter & Value (unit)\\
\hline
$\Gamma_{0,2}$ & $1.19$ (-)\\
$\mathcal{A}$ & $0.819$ (GPa)\\
$\mathcal{B}$ & $1.51$ (GPa)\\
$\mathcal{C}$ & $4.53$ (-)\\
$\mathcal{D}$ & $1.42$ (-)\\
$\rho_{0,2}$ & $1134$ ($\mathrm{kg}$ $\mathrm{m}^{-3}$)\\
$c_{v,2}$ & $1714$ ($\mathrm{J}$ $\mathrm{kg}^{-1}\mathrm{m}^{-3}$)\\
\hline
\end{tabular}
\end{center} 
\end{table}

\begin{table}
\begin{center}
\caption{Parameters for the reaction law of liquid NM}
\label{Tab3}
\begin{tabular}{| c | c |}
\hline
Parameter & Value (unit)\\
\hline
$T_\mathrm{a}$ & $11350$ (K)\\
$C$ & $2.6 \times 10^9$ ($\mathrm{s}^{-1}$)\\
$Q$ & $4.46 \times 10^6$ ($\mathrm{J}$ $\mathrm{kg}^{-1}$)\\
\hline
\end{tabular}
\end{center} 
\end{table}

\subsection{Spatial distribution of cavities}
\label{sec2_2}

\begin{figure*}
\centerline{\includegraphics[width=0.7\textwidth]{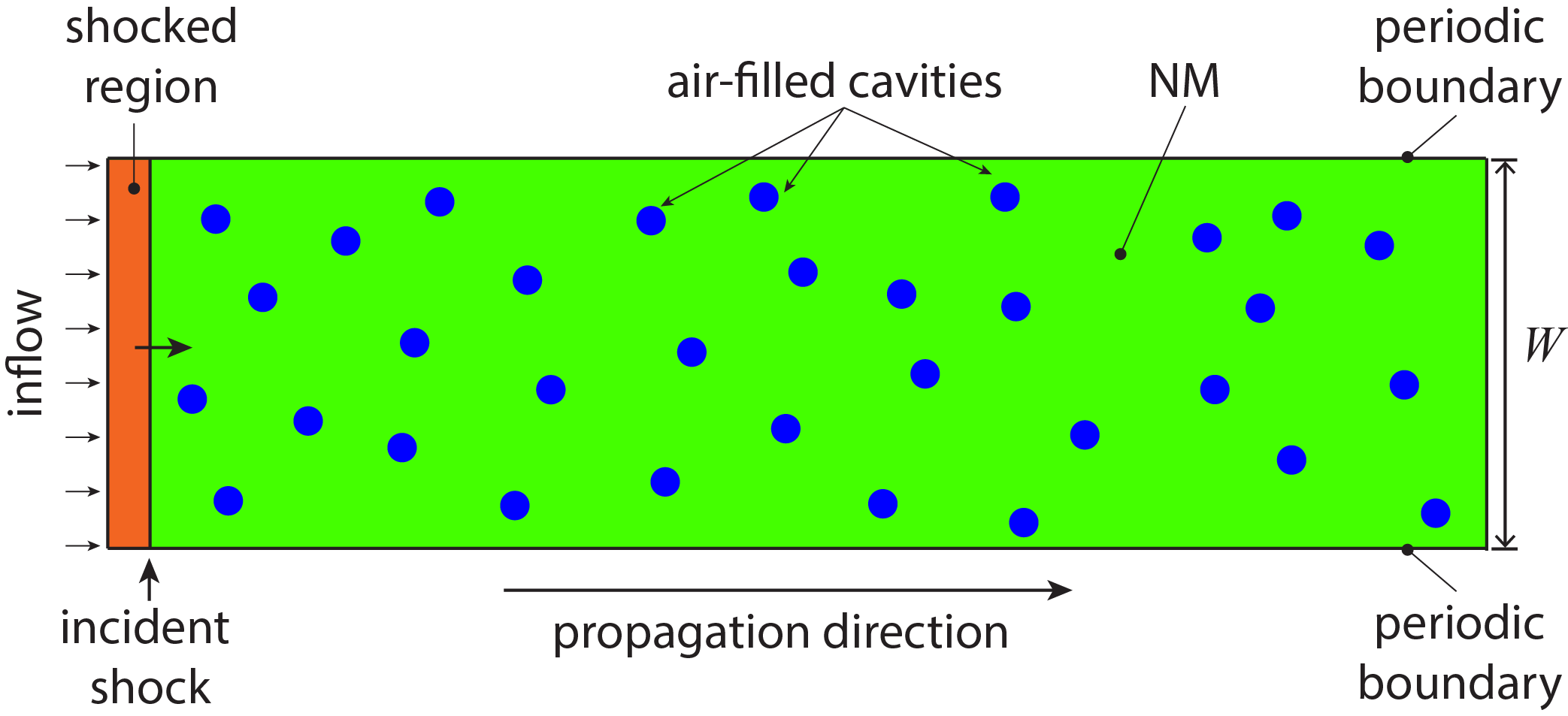}}
		\caption{Schematic illustration of the initial configuration and boundary conditions of the computational domain.}
	\label{problem_config}
\end{figure*}

\begin{figure*}
\centerline{\includegraphics[width=0.8\textwidth]{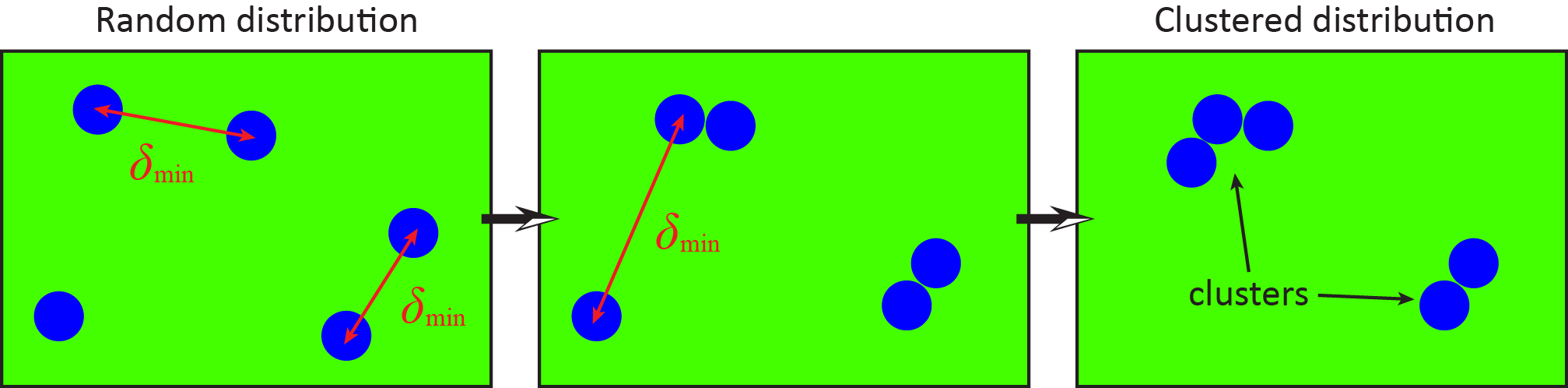}}
		\caption{Schematic illustrating the implementation of clusters of air-filled cavities.}
	\label{clusters}
\end{figure*}

\begin{figure*}
\centerline{\includegraphics[width=0.35\textwidth]{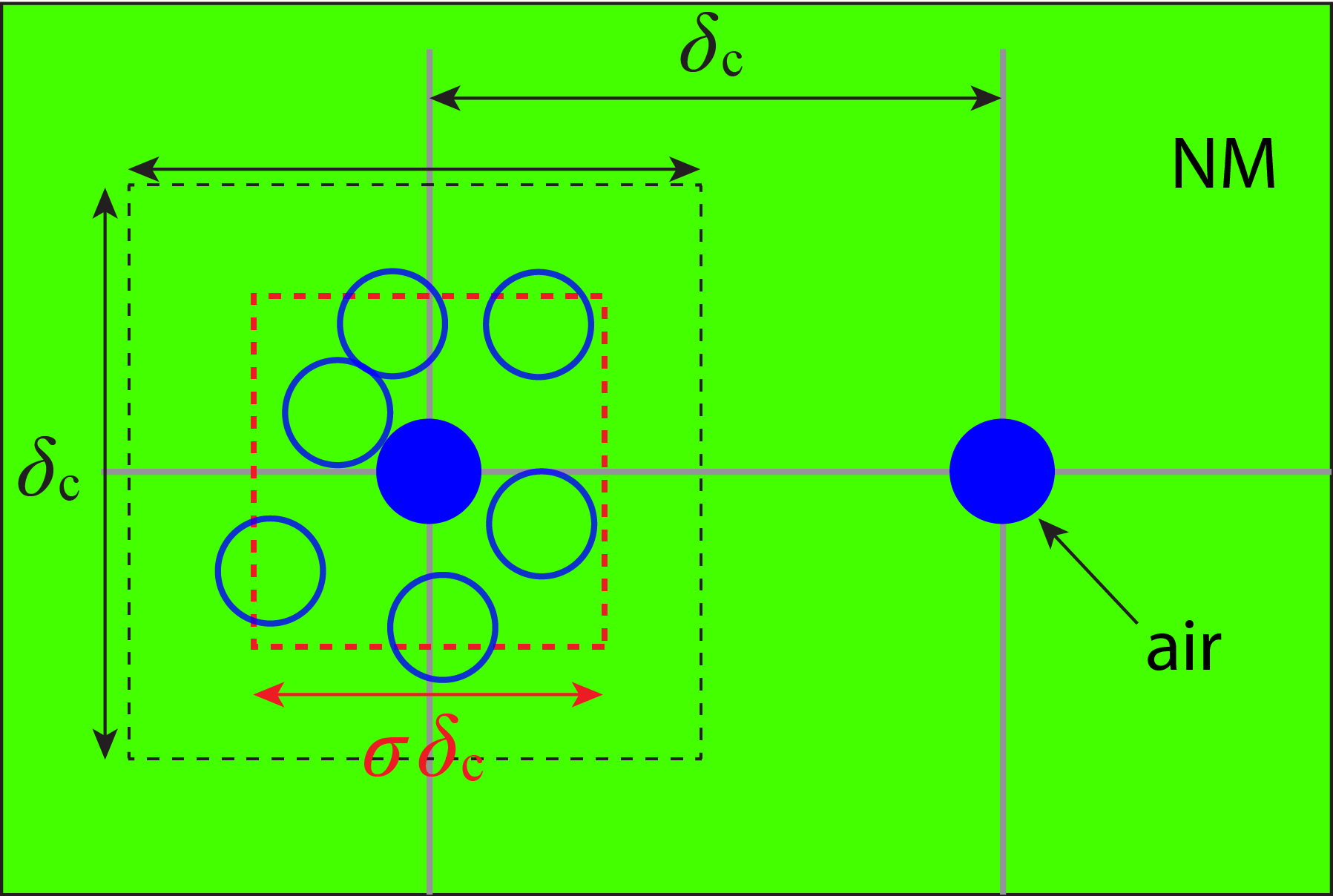}}
		\caption{Schematic showing how the positions of regularly spaced cavities are slightly perturbed.}
	\label{perturbation}
\end{figure*}

\begin{figure*}
\centerline{\includegraphics[width=0.95\textwidth]{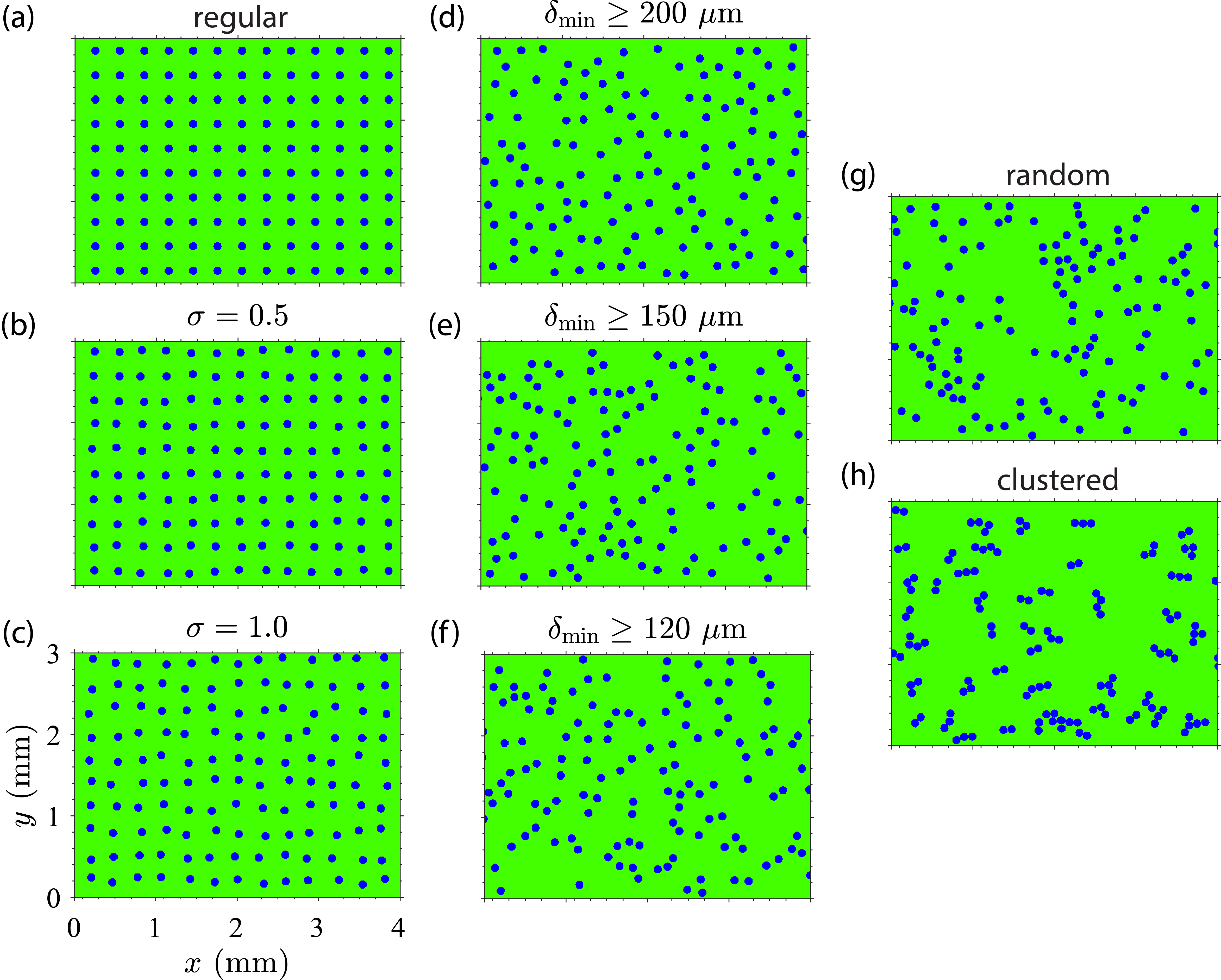}}
		\caption{Sample plots showing various spatial distributions of air-filled cavities of $d_\mathrm{c}=\SI{100}{\micro\metre}$ and $\delta_\mathrm{c}=\SI{300}{\micro\metre}$: (a) A regular array of cavities, slightly perturbed regular arrays of (b) $\sigma=0.5$ and (c) $\sigma=1.0$, uniformly random distributions with an imposed lower limit of (d) $\delta_\mathrm{min} \geq \SI{200}{\micro\meter}$, (e) $\delta_\mathrm{min} \geq \SI{150}{\micro\meter}$, and (f) $\delta_\mathrm{min} \geq \SI{200}{\micro\meter}$, in the minimum spacing among the cavities, (g) a random distribution, and (h) clustered cavities.}
	\label{sample_distribution}
\end{figure*}

The initial configuration and boundary conditions of the computational domain are illustrated in Fig.~\ref{problem_config}. A rightward-moving incident shock wave is introduced to the domain via initially placing a shocked region near the left end of the domain with an inflow (transmissive) boundary condition applied upon the left boundary. The inflow inert material from the left boundary acts as a piston supporting the incident shock wave to propagate into the explosive. Periodic boundary conditions are applied along the top and bottom boundaries. The initial shocked region is $0.5$-millimetre wide (in the $x$-direction) in order to prevent the influence from any transversely moving waves interacting with the left boundary on the explosive medium. Circular-shaped air-filled cavities are considered in this study. In each simulation, the diameter of each cavity $d_\mathrm{c}$ and the average spacing between each two neighbouring cavities $\delta_\mathrm{c}$ are fixed. Note that the spacing between two cavities is measured between the cavity centres. The overall porosity $\phi$ (i.e., volume fraction of air) can be calculated as
\begin{equation}
    \phi = \frac{\pi {d_\mathrm{c}}^2}{4 {\delta_\mathrm{c}}^2}
    \label{Eq8}
\end{equation}
The minimum spacing, i.e., distance between each cavity and its closest neighbour, is denoted as $\delta_\mathrm{min}$. The mean minimum spacing $\overline{\delta}_\mathrm{min}$ of $N$ cavities can be calculated as follows,
\begin{equation}
    \overline{\delta}_\mathrm{min} = \frac{\sum_{i}^{N} \delta_{\mathrm{min},i}}{N}
    \label{Eq9}
\end{equation}
where $i$ is the index of each cavity. Various types of spatial distributions of cavities considered in this study are described as follows:
\begin{description}
\item[\textemdash~Regular distribution] This distribution represents an array of regularly spaced cavities where $\overline{\delta}_\mathrm{min} = \delta_\mathrm{c}$. A sample regular distribution of cavities of $d_\mathrm{c}=\SI{100}{\micro\metre}$ and $\delta_\mathrm{c}=\SI{300}{\micro\metre}$ is shown in Fig.~\ref{sample_distribution}(a).
\item[\textemdash~Random distribution] The positions of the cavities are initialized via a Poisson process while imposing the requirement that the cavities do not overlap, i.e., $\delta_\mathrm{min} \geq d_\mathrm{c}$, as shown in Fig.~\ref{sample_distribution}(g).
\item[\textemdash~Clustered distribution] Firstly, a random distribution of cavities is generated, and then each cavity is moved to its closest neighbour. This clustering process stops once every cavity is clustered with at least another cavity. Fig.~\ref{clusters} illustrates how this clustered distribution is generated and Fig.~\ref{sample_distribution}(h) shows a sample clustered distribution.
\item[\textemdash~Uniformly random distribution] This represents a random distribution of cavities with an imposed lower limit in the spacing between each cavity and its closest neighbour $\delta_\mathrm{min}$. Subfigures (d), (e), and (f) of Fig.~\ref{sample_distribution} are sample plots showing uniformly random distributions of $\delta_\mathrm{min} \geq \SI{200}{\micro\metre}$, $\geq \SI{150}{\micro\metre}$, and $\geq \SI{120}{\micro\metre}$, respectively.
\item[\textemdash~Perturbed regular distribution] This represents a regular distribution with small perturbations introduced to the position of each cavity. As illustrated in Fig.~\ref{perturbation}, the initial position of a cavity is randomly perturbed within a square of box with a side length of $\sigma \delta_\mathrm{c}$ (indicated by the red dashed lines). The parameter $\sigma$ can be considered as the extent of perturbation from a regular array: At $\sigma = 0$, the distribution remains a regular array; in the limit of $\sigma \to \infty$, the distribution approaches a random distribution. In this study, only slightly perturbed cases with $\sigma \leq 1$ are considered. Sample plots of distributions with $\sigma = 0.5$ and $1$ are shown in Fig.~\ref{sample_distribution}(b) and (c), respectively.
\end{description}
Note that, for the cases with a regular distribution, simulations can be performed with one array of cavities, i.e., with a domain width $W=\delta_\mathrm{c}$, and periodic boundary conditions along the top and bottom boundaries. For illustration purposes only, a selected case with a regular distribution (shown in Figs.~\ref{NM_density} and \ref{NM_density_temperature}) was performed with a large domain width $W=\SI{3}{\milli\metre}$.

\section{\label{sec3}Numerical methodology}

The simulation code used to solve the two-dimensional, reactive Euler equations is based upon a uniform Cartesian grid. The MUSCL-Hancock scheme with the van Leer non-smooth slope limiter and a Harten-Lax-van Leer-contact (HLLC) approximate solver for the Riemann problem were used.~\cite{Toro2009} The Strang splitting method was adopted to treat separately the hydrodynamic processes and the reactive processes. This numerical scheme is thus of second-order accuracy in space and time. A diffuse-interface approach~\cite{Michael2016JCP} was used to separate the two immiscible materials, i.e., air and liquid NM. This code is implemented in a CUDA-based (CUDA is an acronym for Compute Unified Device Architecture) parallel-computing framework. The simulations were performed on NVIDIA Tesla V100 16GB general-purpose graphic processing units (GPGPUs). The use of GPU-accelerated computing platforms has been explored in modelling both gaseous~\cite{Mi2017PRF,Mi2018SWJ} and condensed-phase~\cite{Mi2019JAP} detonations. A more detailed description of the herein utilized numerical methodology and its implementation can be found in~\cite{Mi2019JAP}.

\section{\label{sec4}Results and analysis}

For most of the cases in this study, a cavity diameter $d_\mathrm{c}=\SI{100}{\micro\metre}$ and an average spacing $\delta_\mathrm{c}=\SI{300}{\micro\metre}$ are selected, resulting in an overall porosity of $\phi \approx 8.73~\%$. The selected $d_\mathrm{c}$ is close to the range in size of GMBs used in experiments, but slightly greater than the mean values, e.g., $d_\mathrm{c}=\SI{40}{\micro\metre}$ in~\cite{Dattelbaum2010} and $d_\mathrm{c}=\SI{65}{\micro\metre}$ in~\cite{Higgins2018APS}. The selected $\phi$ is much greater than the values found in experimental studies, e.g., $\phi=0.84~\%$ and $2.8~\%$ in~\cite{Dattelbaum2010}. These values of $d_\mathrm{c}$ and $\phi$, which deviate from the experimental parameters, are chosen to ensure a significant sensitising effect in a two-dimensional representation of the heterogeneous explosive mixtures. As shown in the previous study~\cite{Mi2019JAP}, significant hot-spot-driven SDT behaviours---the SDT process in a cavity-laden NM mixture occurs significantly more rapidly than that in neat NM---are observed in a range of relatively low incident shock pressures ($\leq \SI{9}{\giga\pascal}$). Incident shock pressures (denoted as $p_\mathrm{I}$) ranging from $\SI{7.0}{\giga\pascal}$ to $\SI{8.2}{\giga\pascal}$ are thus considered in this study to focus on the hot-spot-driven regime. At each selected $p_\mathrm{I}$, a one-dimensional simulation for the case of neat NM was performed.\\

For the selected cavity diameter $d_\mathrm{c}=\SI{100}{\micro\metre}$, the simulations were performed at a numerical resolution of $\mathrm{d}x=\mathrm{d}y=\SI{1}{\micro\metre}$, ensuring $100$ computational cells across the diameter of a cavity, and approximately $200$ computational cells per reaction-zone length of the Zel'dovich-von Neumann-D\"{o}ring (ZND) detonation profile of neat liquid NM (with the current reaction rate model). As demonstrated in the previous study~\cite{Mi2019JAP}, this numerical resolution is sufficiently fine to obtain a converged result of the characteristic SDT time scale for the herein considered explosive system. For an average cavity spacing $\delta_\mathrm{c}=\SI{300}{\micro\metre}$, the width of the computational domain in the $y$-direction is set to be $W=\SI{3}{\milli\metre}$. For the cases with randomly distributed and clustered cavities, this domain width is not large enough to obtain statistically converged results with one simulation for the range of $p_\mathrm{I}$ considered in this study. Thus, an ensemble of ten simulations were performed for some selected cases with a random or clustered distribution in order to obtain statistically meaningful results.

\subsection{\label{sec4_1}Regular, random, and clustered distributions}

\subsubsection{\label{sec4_1_1}Wave structure}
\begin{figure*}
\centerline{\includegraphics[width=0.65\textwidth]{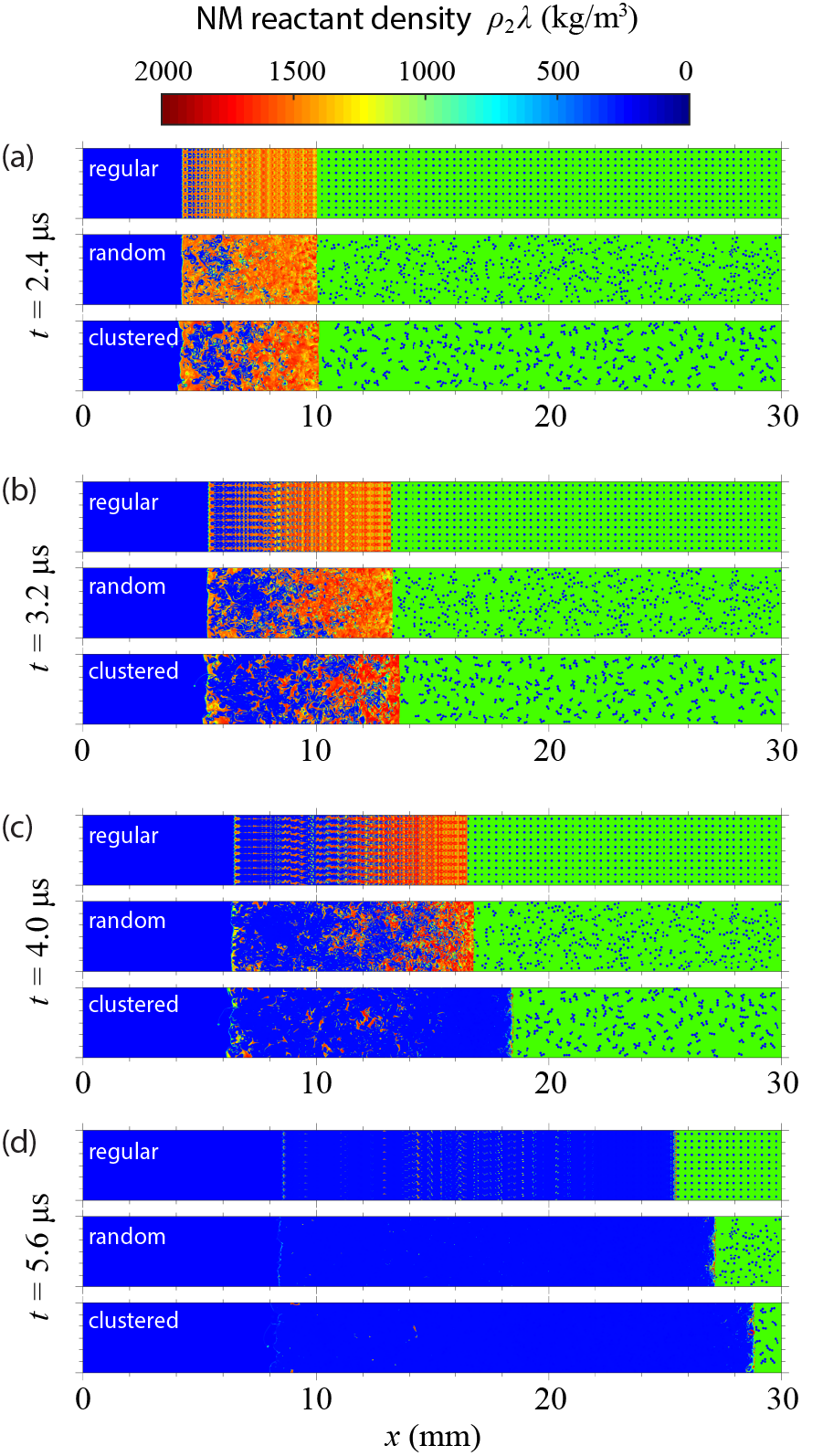}}
		\caption{Colour contour plots of NM reactant density ($\rho_2 \lambda$) showing the evolution of wave structure over the SDT process in mixtures of NM with regularly spaced (top subplot), randomly distributed (middle subplot), and clustered (bottom subplot) air-filled cavities ($d_\mathrm{c}=\SI{100}{\micro\metre}$ and $\delta_\mathrm{c}=\SI{300}{\micro\metre}$) subjected to an incident shock of $\SI{7.0}{\giga\pascal}$ at four different times: (a) $t=\SI{2.4}{\micro\second}$, (b) $t=\SI{3.2}{\micro\second}$, (c) $t=\SI{4.0}{\micro\second}$, and (d) $t=\SI{5.6}{\micro\second}$.}
	\label{NM_density}
\end{figure*}

\begin{figure*}
\centerline{\includegraphics[width=0.65\textwidth]{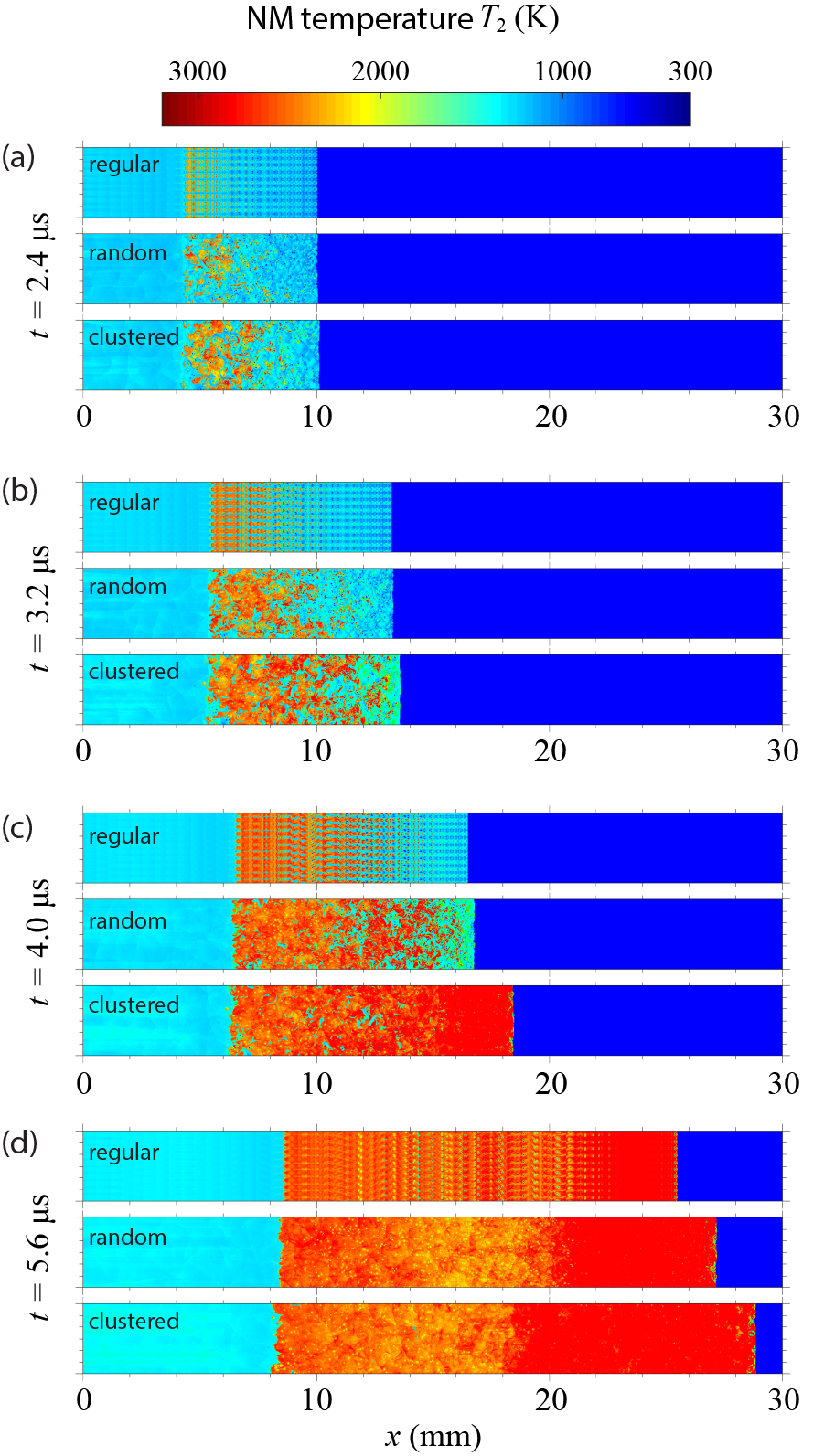}}
		\caption{Colour contour plots of NM temperature ($T_2$) showing the evolution of wave structure over the SDT process in mixtures of NM with regularly spaced (top subplot), randomly distributed (middle subplot), and clustered (bottom subplot) air-filled cavities ($d_\mathrm{c}=\SI{100}{\micro\metre}$ and $\delta_\mathrm{c}=\SI{300}{\micro\metre}$) subjected to an incident shock of $\SI{7.0}{\giga\pascal}$ at four different times: (a) $t=\SI{2.4}{\micro\second}$, (b) $t=\SI{3.2}{\micro\second}$, (c) $t=\SI{4.0}{\micro\second}$, and (d) $t=\SI{5.6}{\micro\second}$.}
	\label{NM_temperature}
\end{figure*}

\begin{figure*}
\centerline{\includegraphics[width=0.65\textwidth]{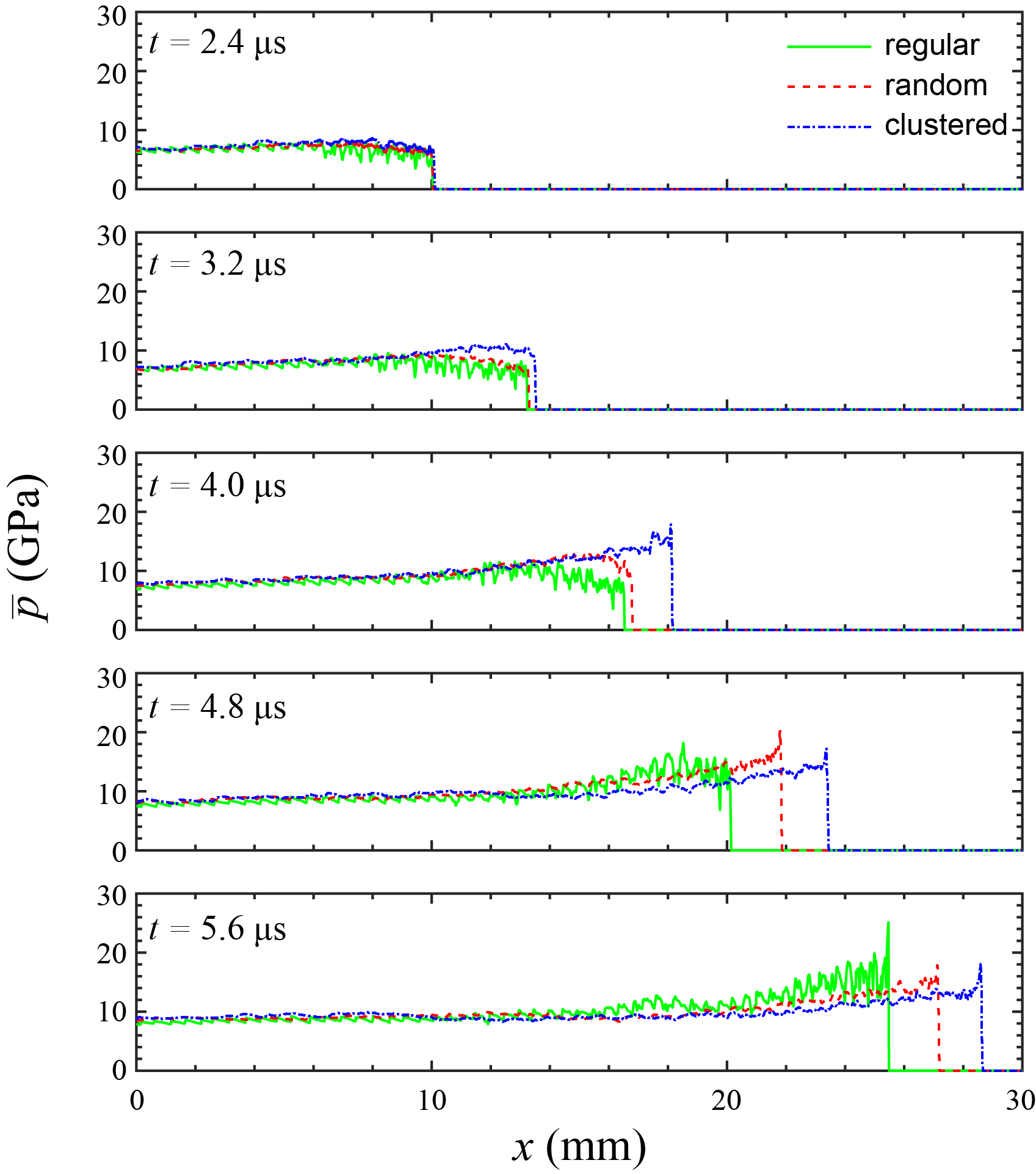}}
		\caption{Profiles of spatially averaged pressure ($\bar{p} (x,t)$) over the domain width $W$ in the $y$-direction at different times throughout the SDT process for the cases with regularly spaced (green curve), randomly distributed (red dash curve), and clustered (blue dash-dot curve) air-filled cavities ($d_\mathrm{c}=\SI{100}{\micro\metre}$ and $\delta_\mathrm{c}=\SI{300}{\micro\metre}$) subjected to an incident shock of $\SI{7.0}{\giga\pascal}$ at four different times.}
	\label{pavg}
\end{figure*}

\begin{figure*}
\centerline{\includegraphics[width=0.8\textwidth]{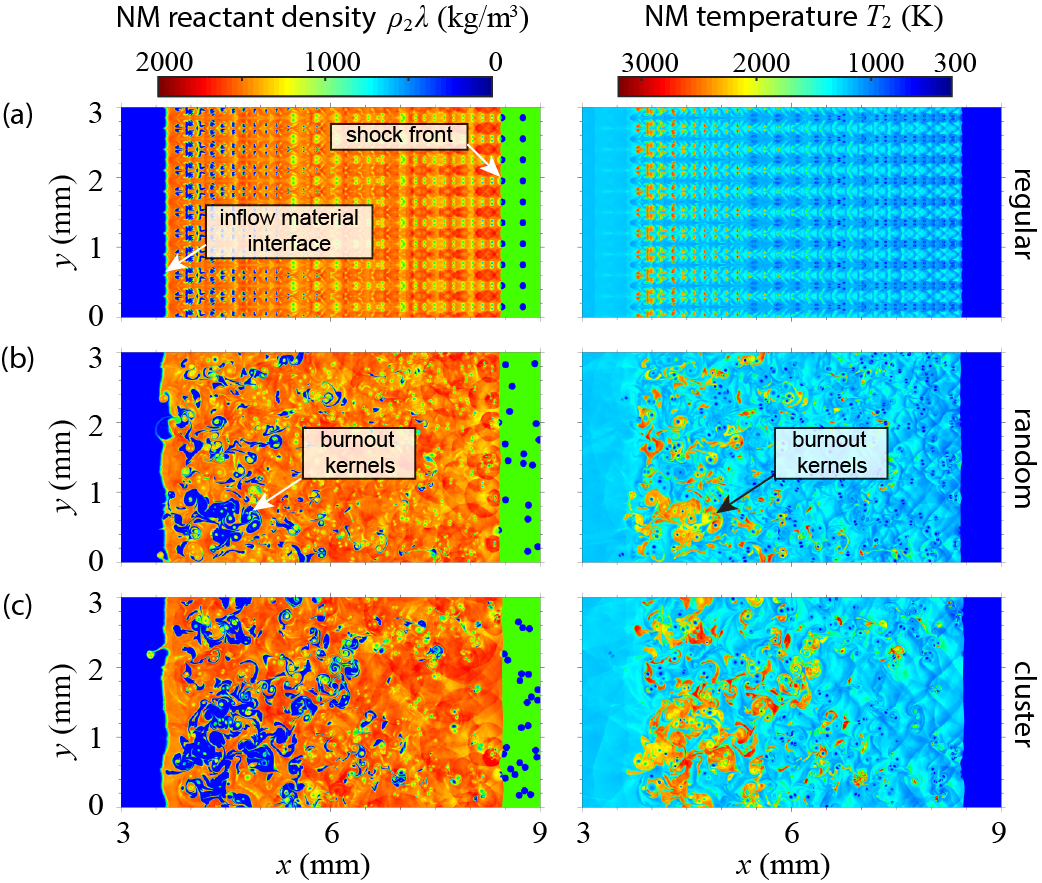}}
		\caption{Colour contour plots of NM reactant density $\rho_2 \lambda$ (left column) and temperature $T_2$ (right column) showing the wave structure at an early stage ($t=\SI{2.0}{\micro\second}$) of the SDT process in mixtures of NM with (a) regularly spaced, (b) randomly distributed, and (c) clustered air-filled cavities ($d_\mathrm{c}=\SI{100}{\micro\metre}$, $\delta_\mathrm{c}=\SI{300}{\micro\metre}$) subjected to an incident shock of $\SI{7.0}{\giga\pascal}$.}
	\label{NM_density_temperature}
\end{figure*}

The simulation results for the cases with regularly spaced, randomly distributed, and clustered cavities are first compared in this subsection. In Figs.~\ref{NM_density} and~\ref{NM_temperature}, colour contour plots of NM reactant density ($\rho_2 \lambda$) and temperature ($T_2$), respectively, show the evolving flow fields resulting from the cases with regularly spaced (top), randomly distributed (middle), and clustered (bottom) cavities subjected to an incident shock of $p_\mathrm{I}=\SI{7.0}{\giga\pascal}$ at four different times throughout the SDT process. A rightward-propagating shock front can be identified in all of these contour plots. In the plots of NM reactant density (Fig.~\ref{NM_density}), the blue regions near the left end of the domain, which can be more clearly seen at early times, indicate the inflow of an inert ``piston'' material following the incident shock.  A red-orange-coloured region between the shock front and the inflow material interface is the reaction zone of shocked material. The regions wherein NM is fully reacted, i.e., $\lambda = 0$, appear to be blue between the shock front and the inflow material interface. These aforementioned features can be more clearly observed in the zoom-in views (provided in Fig.~\ref{NM_density_temperature}) of the NM reactant density and temperature fields in the shock-induced reaction zone at an early time $t=\SI{2.0}{\micro\second}$. Burnout regions with temperatures above $\SI{2000}{K}$ are increasingly more populated from the case with a regular distribution to the case with clustered cavities. As shown in Fig.~\ref{NM_density}(a)-(c), the burnout regions first appear in the reaction zone near the inflow material interface, and gradually catch up with the leading shock front. At a late time $t=\SI{5.6}{\micro\second}$ as shown in Fig.~\ref{NM_density}(d), the incident shock wave has evolved into a detonation wave wherein the reaction zone is very thin and closely attached to the shock front. A detonation wave can be identified from the temperature plots shown in Fig.~\ref{NM_temperature}(c) and (d) as a region of high temperature ($\approx \SI{3000}{\kelvin}$) immediately attached to the leading shock front. The resulting wave structure evolves in an increasingly faster pace from the case with a regular array of cavities to the case with a clustered distribution. In addition to the qualitative information revealed by the contour plots of the flow field (Figs.~\ref{NM_density} and~\ref{NM_temperature}), profiles of spatially averaged pressure ($\bar{p} (x,t)$) over the domain width $W$ in the $y$-direction at different times throughout the SDT process for the three corresponding cases with regularly spaced (green curve), randomly distributed (red dash curve), and clustered (blue dash-dot curve) air-filled cavities are shown in Fig.~\ref{pavg}. The spatially averaged pressure is calculated as follows,
\begin{equation}
\bar{p}(x,t) = \frac{1}{W} \int_{0}^{W} p(x,y,t) \mathrm{d}y
\end{equation}
A gradual increase in pressure immediately behind the leading shock wave is observed for all of the cases with different cavity distributions. Quantitative results of peak pressure can be extracted from these plots for future studies.\\

\subsubsection{\label{sec4_1_2}Detonation overtake time and Pop-plot}
\begin{figure*}
\centerline{\includegraphics[width=0.9\textwidth]{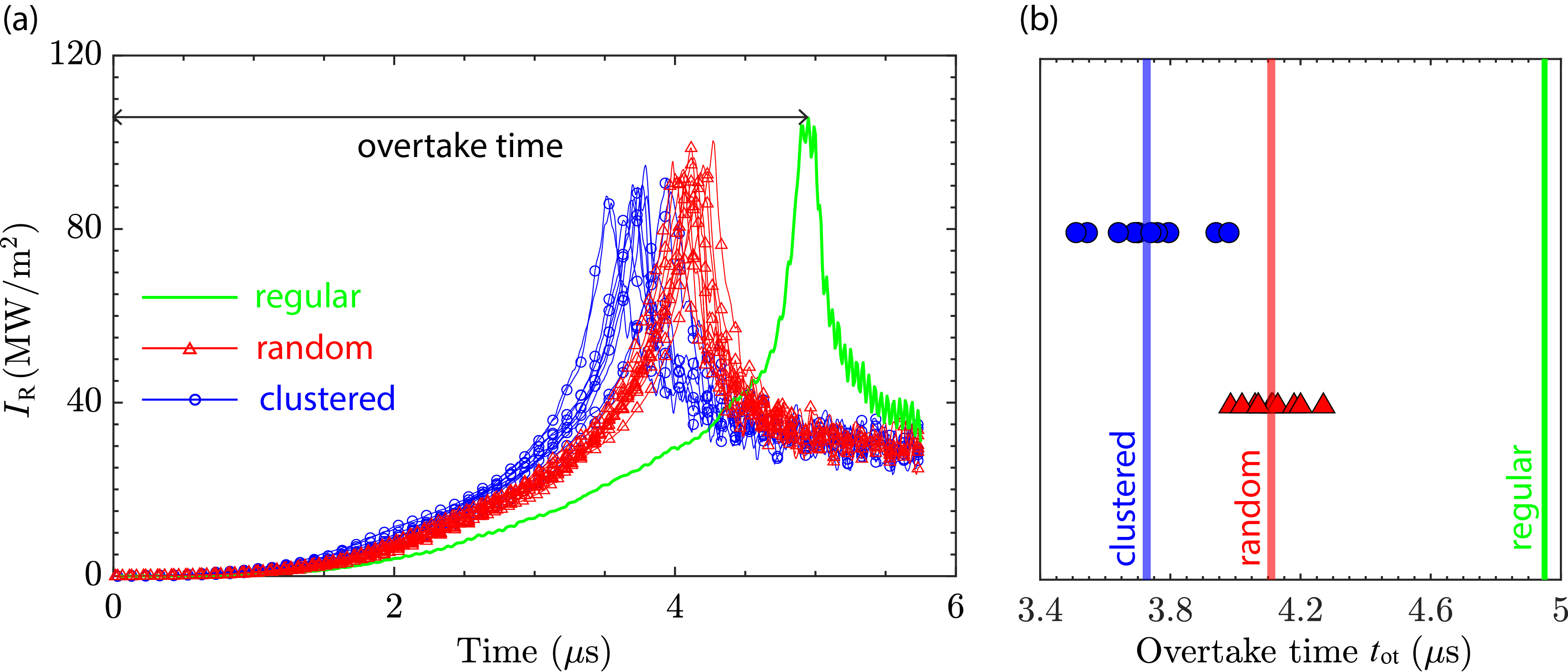}}
		\caption{(a) Specific rate of energy release per unit cross-sectional area of the explosive $I_\mathrm{R}$ plotted as a function time for an incident shock of $p_\mathrm{I}=\SI{7.0}{\giga\pascal}$ with regularly spaced (green curve), randomly distributed (red curves with triangles), and clustered (blue curves with circles) air-filled cavities ($d_\mathrm{c}=\SI{100}{\micro\metre}$, $\delta_\mathrm{c}=\SI{300}{\micro\metre}$), showing that detonation overtake time $t_\mathrm{ot}$ is measured as the time associated with the maximum $I_\mathrm{R}$. For the cases with random and clustered distributions, the results of an ensemble of ten simulations are plotted. (b) A scatter chart showing $t_\mathrm{ot}$ resulting from the ensemble simulations for the cases with randomly distributed (red triangles) and clustered (blue circles) cavities.}
	\label{reaction_rate}
\end{figure*}

To quantitatively examine how fast the SDT process occurs in each scenario of cavity distribution, the time history of the specific rate of energy release per unit cross-sectional area of the explosive $I_\mathrm{R}(t)$ ($\sim \SI{}{\watt\per\metre\squared}$) is obtained for each simulation. At any time $t$, $I_\mathrm{R}$ of the entire reacting system can be calculated as follows,
\begin{equation}
    I_\mathrm{R}(t) = \frac{1}{W} \int_{0}^{W} \int_{0}^{L} \dot{q} (x,y,t) \mathrm{d}x \mathrm{d}y
\end{equation}
where $L$ and $W$ are the length and width of the domain in the $x$- and $y$-directions, respectively, and $\dot{q} = -Q z_2 \rho_2 K$ is the volumetric rate of energy release ($\sim \SI{}{\watt\per\cubic\metre}$) at each computational cell of the domain. The result of $I_\mathrm{R}(t)$ for the case of a regular array of cavities subjected to an incident shock pressure of $\SI{7.0}{\giga\pascal}$ is plotted as the green curve in Fig.~\ref{reaction_rate}(a). The ensembles of results for the cases with randomly distributed and clustered cavities are plotted in Fig.~\ref{reaction_rate}(a) as red curves with triangle markers and blue curves with circle markers, respectively. Each curve represents the $I_\mathrm{R}(t)$ resulting from one simulation within the ensemble.\\

For all of the cases subjected to an incident shock pressure of $\SI{7.0}{\giga\pascal}$, as shown in Fig.~\ref{reaction_rate}, the rate of energy release increases gradually with time and, after reaching a maximum, decreases to a quasi-steady value. As verified in the previous study~\cite{Mi2019JAP}, the time at which  $I_\mathrm{R}$ reaches its peak value was associated with the maximum acceleration of the leading shock in the SDT process. The time of maximum shock acceleration is commonly measured as the characteristic time of the SDT process, i.e., detonation overtake time, in experimental studies of heterogeneous explosives. Thus, in this study, the detonation overtake time (denoted as $t_\mathrm{ot}$) is determined as the time of maximum $I_\mathrm{R}$.\\

The results of overtake times $t_\mathrm{ot}$ for cases shown in Fig.~\ref{reaction_rate}(a) are summarized as a scatter chart in Fig.~\ref{reaction_rate}(b). The green vertical line is the $t_\mathrm{ot}$ resulting from the case with regularly spaced cavities. The ensembles of results for the cases with a random distribution and a clustered distribution are plotted as red triangles and blue circles, respectively. The vertical line indicates the mean $t_\mathrm{ot}$ of each ensemble. The mean $t_\mathrm{ot}$ for randomly distributed and clustered cavities are significantly shorter than that for a regular array. Despite a small overlap, most of the overtake times resulting from clustered distributions are less than those resulting from random distributions.\\ 

\begin{figure*}
\centerline{\includegraphics[width=0.75\textwidth]{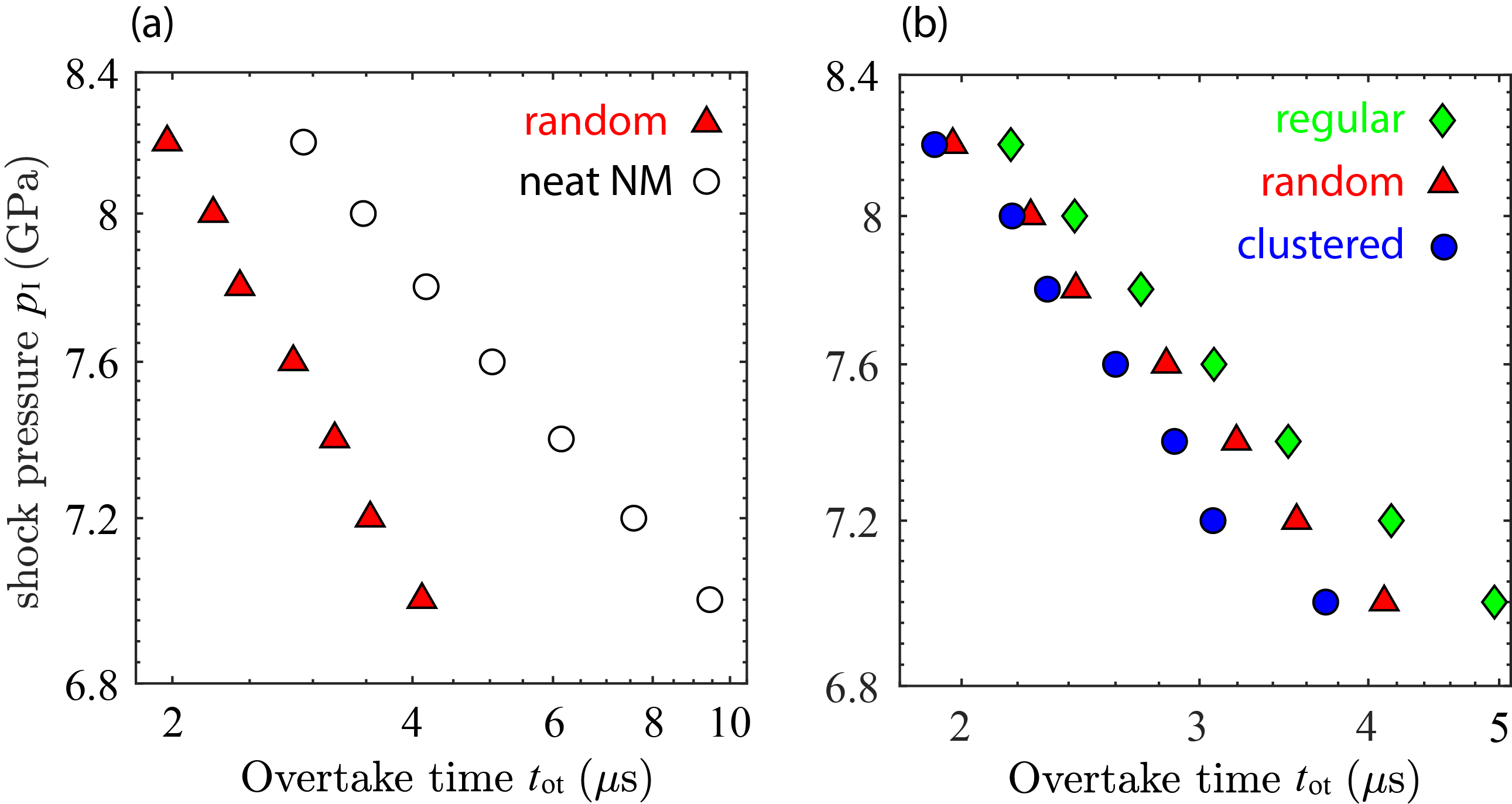}}
		\caption{Simulation results of overtake time $t_\mathrm{ot}$ as a function of incident shock pressure are presented as a Pop-plot (i.e., with log-log scales). The results for the cases with neat NM (open black circles) and randomly distributed cavities ($d_\mathrm{c}=\SI{100}{\micro\metre}$, $\delta_\mathrm{c}=\SI{300}{\micro\metre}$) are compared in (a); the results for the cases with regularly spaced (green diamonds), randomly distributed (red triangles), and clustered (solid blue circles) air-filled cavities are shown in (b). Note that the data for random and clustered distributions at $p_\mathrm{I}=\SI{7.0}{\giga\pascal}$ are the averaged overtake time based on an ensemble of ten simulations. The rest of the data are the results of individual simulations.}
	\label{Pop_plot}
\end{figure*}

Since the ensembles of results at an incident shock pressure of $\SI{7.0}{\giga\pascal}$ show that a clustered distribution of cavities consistently results in a smaller $t_\mathrm{ot}$ than that from a random distribution, only one simulation for these two types of cavity distributions were performed for the cases with shock pressures ranging from $\SI{7.2}{\giga\pascal}$ to $\SI{8.2}{\giga\pascal}$. The results of $t_\mathrm{ot}$ are plotted as a function of incident shock pressure, i.e., a Pop-plot, in Fig.~\ref{Pop_plot}.  As shown in Fig.~\ref{Pop_plot}(a), the $t_\mathrm{ot}$ for the cases with randomly distributed cavities of $d_\mathrm{c}=\SI{100}{\micro\metre}$ and $\delta_\mathrm{c}=\SI{300}{\micro\metre}$ (red triangles) are significantly smaller than those for the cases with neat NM (open black circles).  Figure~\ref{Pop_plot}(b) shows that, throughout the considered range in $p_\mathrm{I}$, the rank-order in $t_\mathrm{ot}$ of regular (green diamonds), random (red triangles), and clustered (solid blue circles) distributions of cavities is consistently the same as that found at $p_\mathrm{I}=\SI{7.0}{\giga\pascal}$ (reported in Fig.~\ref{reaction_rate}).

\subsubsection{\label{sec4_1_3}Statistical analysis}
\begin{figure*}
\centerline{\includegraphics[width=0.75\textwidth]{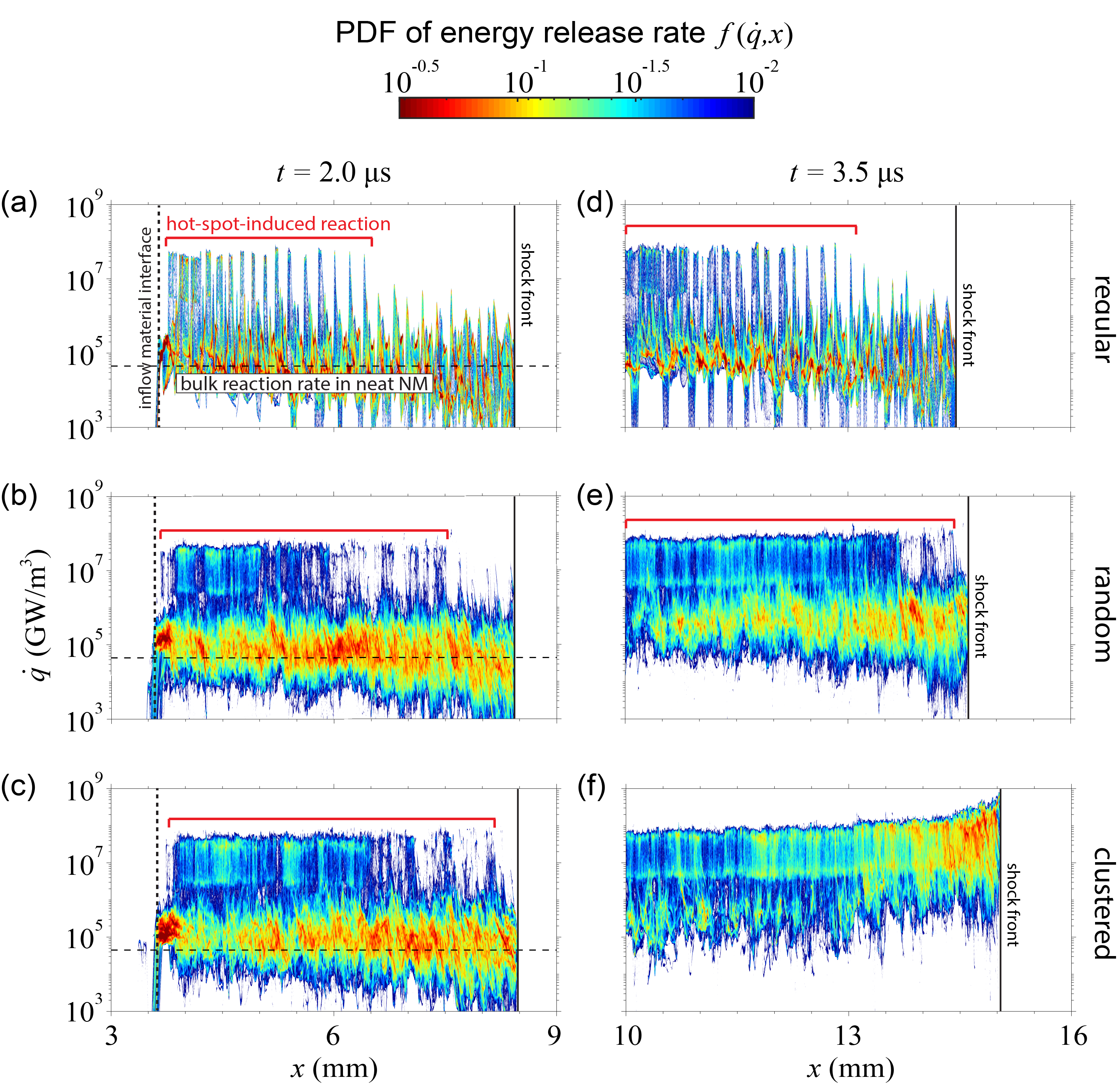}}
		\caption{Probability density function $f(\dot{q},x)$ of volumetric rate of energy release $\dot{q}$ averaged over the width of the domain in the $y$-direction plotted as a function of $x$ for the cases with (top row) regularly spaced, (middle row) randomly distributed, and (bottom row) clustered air-filled cavities ($d_\mathrm{c}=\SI{100}{\micro\metre}$, $\delta_\mathrm{c}=\SI{300}{\micro\metre}$) subjected to an incident shock of $\SI{7.0}{\giga\pascal}$ (a)-(c) at an early time $t=\SI{2.0}{\micro\second}$ (corresponding to the snapshots of wave structures shown in Fig.~\ref{NM_density_temperature}) and (d)-(f) at a later time $t=\SI{3.5}{\micro\second}$.}
	\label{pdf_x}
\end{figure*}

A set of statistical analyses was performed to monitor the rate of energy release induced by the incident shock interacting with cavities throughout the SDT processes. Figure~\ref{pdf_x} shows the probability density function (PDF) of the volumetric rate of energy release as a function of the $x$-coordinate, i.e., $f(\dot{q},x)$, corresponding to the three snapshots presented in Fig.~\ref{NM_density_temperature} at an early state $t=\SI{2.0}{\micro\second}$ of the SDT process and a later time $t=\SI{3.5}{\micro\second}$. In Fig.~\ref{pdf_x}, the vertical axis of a logarithmic scale represents the volumetric rate of energy release $\dot{q}$, and the horizontal axis is the spatial coordinate $x$. The plots at an early time $t=\SI{2.0}{\micro\second}$ shown in Fig.~\ref{pdf_x}(a)-(c) have the same range in $x$-coordinate as those of the zoom-in plots of flow field shown in Fig.~\ref{NM_density_temperature}. The vertical solid line indicates the averaged position of the leading shock front. The vertical dashed line marks the inflow material interface. The horizontal dashed line in Fig.~\ref{pdf_x}(a)-(c) indicates the bulk reaction rate in neat NM subjected to an incident shock of the same strength ($p_\mathrm{I} = \SI{7.0}{\giga\pascal}$). A significant distribution of $f(\dot{q},x)$ over reaction rates that are more than two orders of magnitude greater than the bulk reaction rate can be seen in all of the three cases. The region (along the $x$-direction) with a significant distribution of PDF over $10^7 \, \mathrm{GW}/\mathrm{m^3}$ is marked by the red brackets in Fig.~\ref{pdf_x}. \\

In the case with a regular array of cavities (Fig.~\ref{pdf_x}(a)), only discrete, spike-like distribution patterns of $f(\dot{q},x)$ appear in the range of $\dot{q} \geq 10^7 \, \mathrm{GW}/\mathrm{m^3}$, and this region of fast reaction is significantly behind the leading shock front. In the case with randomly distributed cavities, as shown in Fig.~\ref{pdf_x}(b), the distribution pattern of $f(\dot{q},x)$ above $\dot{q} = 10^7 \, \mathrm{GW}/\mathrm{m^3}$ is more continuous in comparison to the spike-like patterns for the case of a regular distribution; the region of fast reaction, as indicated by the red bracket, is longer and closer to the shock front. In the case with clustered cavities shown in Fig.~\ref{pdf_x}(c), the region of $\dot{q} \geq 10^7 \, \mathrm{GW}/\mathrm{m^3}$ is even larger and closer to the shock front than that resulting from a random distribution.\\

At a later time $t=\SI{3.5}{\micro\second}$, as shown in Fig.~\ref{pdf_x}(f), the region of fast reaction has reached the shock front for the case with clustered cavities. For the case with randomly distributed cavities shown in Fig.~\ref{pdf_x}(e), the region of fast reaction has significantly advanced towards the shock front. For the case with a regular distribution, Fig.~\ref{pdf_x}(d) indicates that the region of fast reaction has slightly advanced, but it is still significantly apart from the shock front.\\

\begin{figure*}
\centerline{\includegraphics[width=0.925\textwidth]{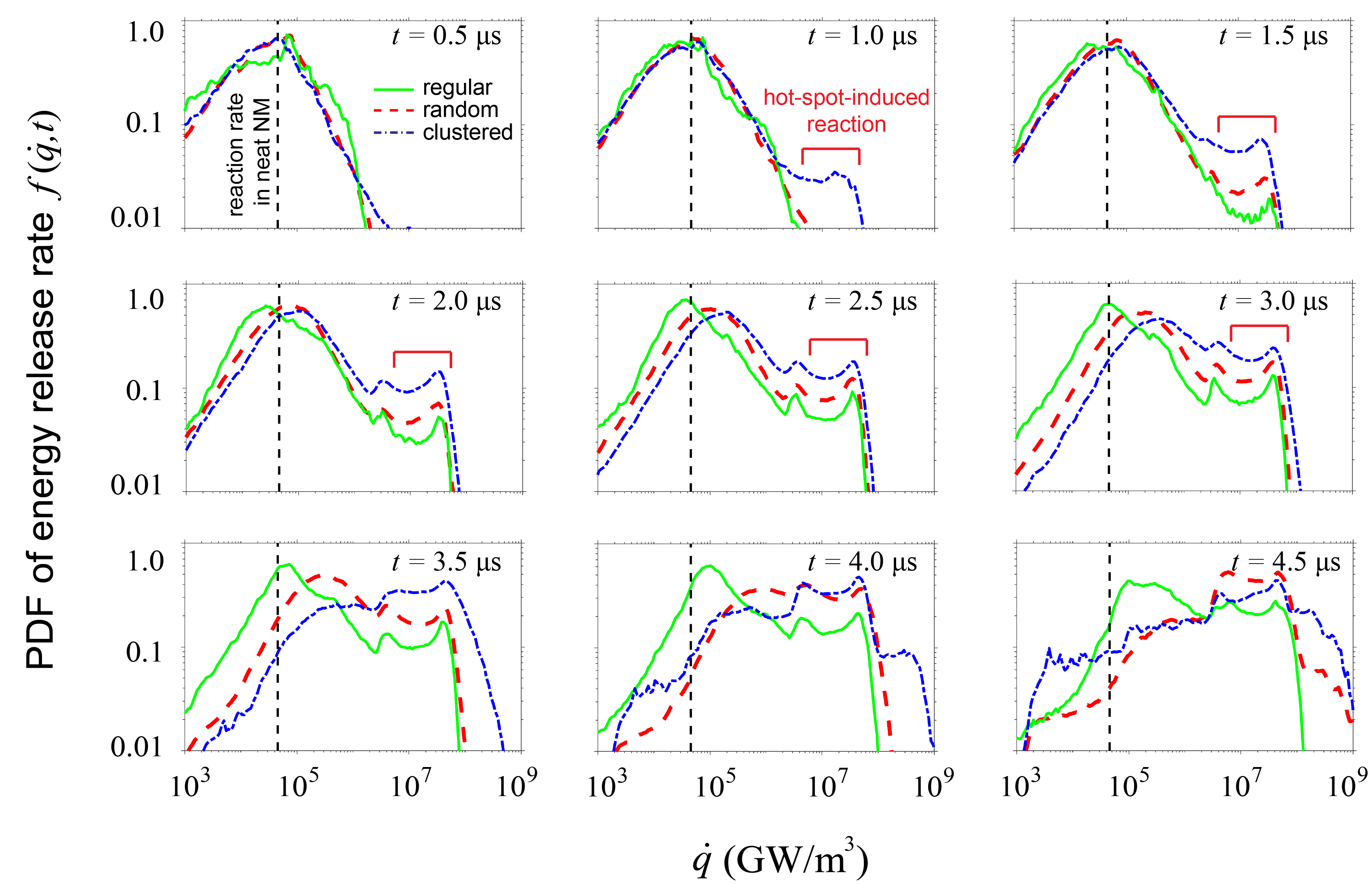}}
		\caption{Probability density function $f(\dot{q},t)$) of volumetric rate of energy release $\dot{q}$ for the entire domain at different times throughout the SDT process for the cases with regularly spaced (green curve), randomly distributed (red dash curve), and clustered (blue dash-dot curve) air-filled cavities ($d_\mathrm{c}=\SI{100}{\micro\metre}$, $\delta_\mathrm{c}=\SI{300}{\micro\metre}$) subjected to an incident shock of $\SI{7.0}{\giga\pascal}$.}
	\label{pdf_t}
\end{figure*}

The PDF of the energy release rate of the entire explosive system as a function of time can be calculated as follows,
\begin{equation}
    f(\dot{q},t) = \int_{0}^{L} f(\dot{q},x,t) \mathrm{d}x
\end{equation}
Figure~\ref{pdf_t} demonstrates the PDF of the reaction rate at different times throughout the complete SDT process resulting from the cases with a regular (green curve), random (red dashed curve), and clustered (blue dot-dashed curve) distribution of cavities at $p_\mathrm{I} = \SI{7.0}{\giga\pascal}$. The peak of the PDF curve at each time roughly indicates the bulk reaction rate of the explosive mixture. The vertical dashed line indicates the bulk reaction rate in neat NM subjected to the same incident shock pressure. The red bracket in the plots from $t=\SI{1.0}{\micro\second}$ to $t=\SI{3.0}{\micro\second}$ indicates the distribution of the PDF around $\dot{q} = 10^7 \, \mathrm{GW}/\mathrm{m^3}$, which is nearly two orders of magnitude faster than the bulk reaction rate. At a very early time $t=\SI{0.5}{\micro\second}$, the PDF of all of the three scenarios peaks at approximately the bulk reaction rate of neat NM. From $t=\SI{1.0}{\micro\second}$ onward, a ``hump'' in the PDF for the cases with a random and clustered distribution arises around $\dot{q} = 10^7 \, \mathrm{GW}/\mathrm{m^3}$; for the case with a regular distribution, a hump at large $\dot{q}$ starts to appear at $t=\SI{1.5}{\micro\second}$. This hump at large $\dot{q}$ grows faster in the case with a clustered distribution than that in the case with a random distribution. The growth of fast reaction in the case with a regular distribution is much slower than that in the other two cases. From $t=\SI{3.0}{\micro\second}$ onward, the peak value of the PDF for both random and clustered distributions moves away from the bulk reaction rate of neat NM and gradually merges into the hump over a range of $\dot{q} \geq 10^7 \, \mathrm{GW}/\mathrm{m^3}$. In the case with a regular distribution, this shift of maximum PDF value towards large $\dot{q}$ occurs significantly later, i.e., after approximately $t=\SI{4.0}{\micro\second}$.\\

The simulations reported in the current section show that a random distribution of cavities results in a smaller detonation overtake time than that resulting from a regular distribution; clustering of randomly distributed cavities further reduces $t_\mathrm{ot}$. The effects of randomness and clustering in the distribution of mesoscale heterogeneities are separately investigated in Sect.~\ref{sec4_2} and Sect.~\ref{sec4_3}, respectively.\\

\subsection{\label{sec4_2}Uniformly random and slightly perturbed regular distributions}
To further examine the effect of randomness in the cavity distribution on the SDT behaviour, additional simulations were performed with two transitional scenarios between a regular array and a purely random distribution of cavities: (1) Uniformly random distribution with an imposed lower limit in $\delta_\mathrm{min}$ (i.e., the distance from a cavity to its nearest neighbour); (2) a slightly perturbed regular distribution.\\

\begin{figure*}
\centerline{\includegraphics[width=0.95\textwidth]{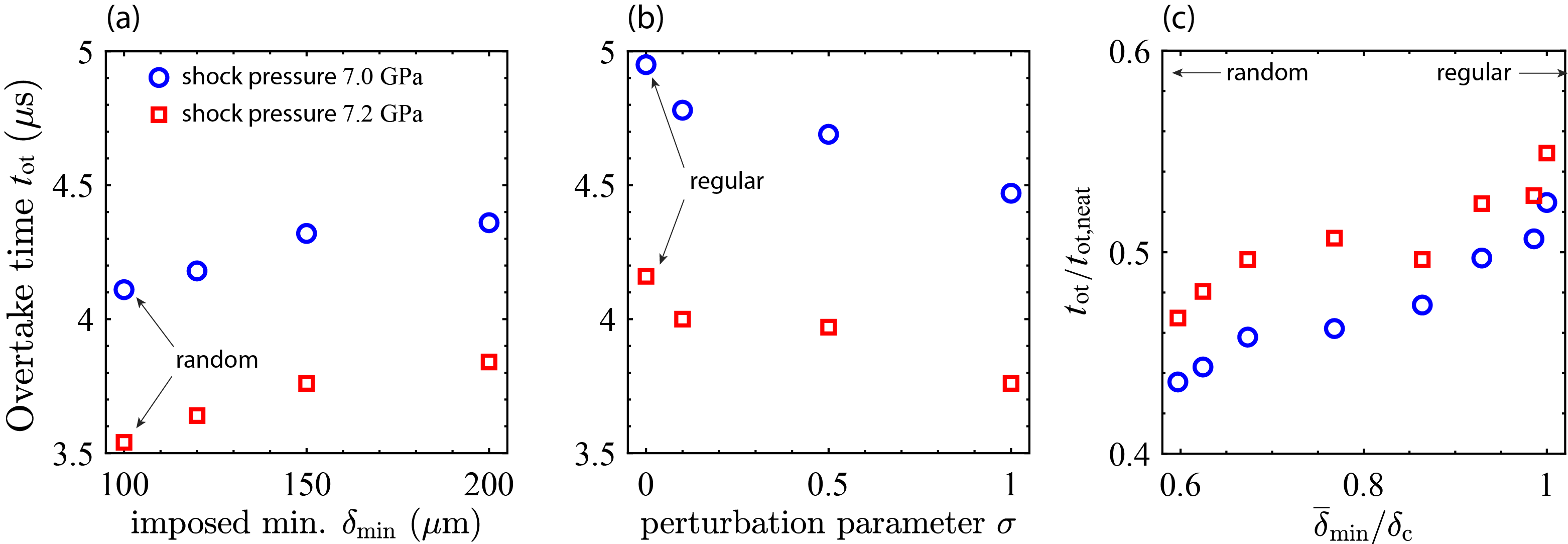}}
		\caption{Detonation overtake times $t_\mathrm{ot}$ resulting from the scenarios of (a) randomly distributed cavities ($d_\mathrm{c}=\SI{100}{\micro\metre}$) with an imposed lower limit in the minimum spacing between cavities (i.e., $\delta_\mathrm{min} \geq$ limit) and (b) regularly spaced cavities with small perturbations subjected to incident shocks of $\SI{7.0}{\giga\pascal}$ (blue circles) and $\SI{7.2}{\giga\pascal}$ (red squares). (c) The results of $t_\mathrm{ot}$ normalized by the overtake time for neat NM subjected to the corresponding shock pressure ($t_\mathrm{ot,neat}$) for both of the scenarios (a) and (b) plotted as a function of the mean minimum spacing $\overline{\delta}_\mathrm{min}$ of each type of distribution normalized by the averaged spacing $\delta_\mathrm{c}=\SI{300}{\micro\metre}$.}
	\label{tot_rmin_pert}
\end{figure*}

For a random distribution of cavities, cavity diameter $d_\mathrm{c}=\SI{100}{\micro\metre}$ is imposed as a lower limit in $\delta_\mathrm{min}$ of the initial cavity positions generated via a Poisson process in order to prevent overlapping. Further increasing this imposed limit in $\delta_\mathrm{min}$, the random distribution becomes more uniform as shown in Fig.~\ref{sample_distribution}(d)-(f). Simulations for the cases with a uniformly random distribution of $\delta_\mathrm{min} \geq \SI{120}{\micro\metre}$, $\geq \SI{150}{\micro\metre}$, and $\geq \SI{200}{\micro\metre}$ were performed. The results of $t_\mathrm{ot}$ (along with that for the case with a random distribution) are plotted as a function of the imposed minimum in $\delta_\mathrm{min}$ in Fig.~\ref{tot_rmin_pert}(a). For both of the shock pressures considered ($p_\mathrm{I}=\SI{7.0}{\giga\pascal}$ and $\SI{7.2}{\giga\pascal}$), $t_\mathrm{ot}$ increases as the imposed minimum spacing increases.\\

In the cases with a slightly perturbed regular array of cavities, the parameter $\sigma$ quantifies the amplitude of perturbation as described in Sect.~\ref{sec2_2}. When $\sigma=0$, the regular array is unperturbed. The value of $\sigma$ indicates the randomness of the spatial distribution of cavities. The resulting $t_\mathrm{ot}$ is plotted as a function of $\sigma$ in Fig.~\ref{tot_rmin_pert}(b). For both of the shock pressures considered ($p_\mathrm{I}=\SI{7.0}{\giga\pascal}$ and $\SI{7.2}{\giga\pascal}$), $t_\mathrm{ot}$ decreases as the $\sigma$ increases.\\

The results of $t_\mathrm{ot}$ for the cases with uniformly random distributions and perturbed regular distributions normalized by the overtake time for the case of neat NM $t_\mathrm{ot,neat}$ are plotted on the same graph (Fig.~\ref{tot_rmin_pert}(c)) as a function of mean distance to the nearest neighbouring cavity $\overline{\delta}_\mathrm{min}$ normalized by the mean cavity spacing $\delta_\mathrm{c}$. For a regular distribution, $\overline{\delta}_\mathrm{min}$ is equal to $\delta_\mathrm{c}$. An increase in the randomness of the cavity distribution is associated with a decrease in $\overline{\delta}_\mathrm{min}$. As shown in Fig.~\ref{tot_rmin_pert}(c), the resulting $t_\mathrm{ot}/t_\mathrm{ot,neat}$ decreases as $\overline{\delta}_\mathrm{min}/\delta_\mathrm{c}$ decreases; the two sets of results for $p_\mathrm{I}=\SI{7.0}{\giga\pascal}$ and $\SI{7.2}{\giga\pascal}$ roughly collapse onto one curve.

\subsection{\label{sec4_3}Comparison between clusters and large cavities}
 
\begin{figure*}
\centerline{\includegraphics[width=0.8\textwidth]{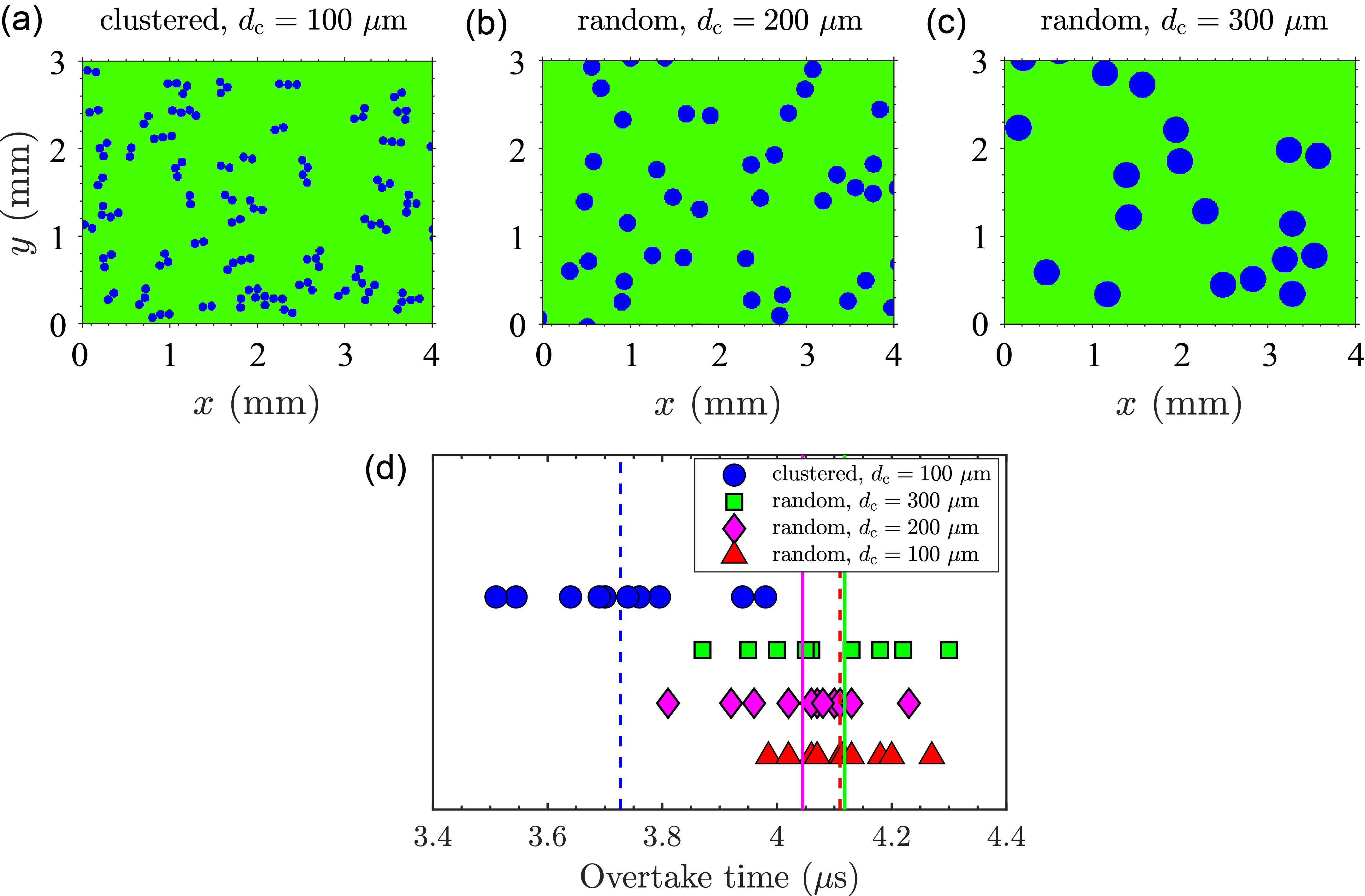}}
		\caption{Sample plots showing (a) clustered cavities of $d_\mathrm{c}=\SI{100}{\micro\metre}$, $\delta_\mathrm{c}=\SI{300}{\micro\metre}$, (b) randomly distributed cavities of $d_\mathrm{c}=\SI{200}{\micro\metre}$, $\delta_\mathrm{c}=\SI{600}{\micro\metre}$, and (c) randomly distributed cavities of $d_\mathrm{c}=\SI{300}{\micro\metre}$, $\delta_\mathrm{c}=\SI{900}{\micro\metre}$. (d) A scatter chart showing the overtake times $t_\mathrm{ot}$ resulting from the ensembles of simulations for the cases with clustered cavities of $d_\mathrm{c}=\SI{100}{\micro\metre}$, $\delta_\mathrm{c}=\SI{300}{\micro\metre}$ (blue circles), randomly distributed cavities of $d_\mathrm{c}=\SI{300}{\micro\metre}$, $\delta_\mathrm{c}=\SI{900}{\micro\metre}$ (green squares), randomly distributed cavities of $d_\mathrm{c}=\SI{200}{\micro\metre}$, $\delta_\mathrm{c}=\SI{600}{\micro\metre}$ (magenta diamonds), and $d_\mathrm{c}=\SI{100}{\micro\metre}$, $\delta_\mathrm{c}=\SI{300}{\micro\metre}$ (red triangles). Each vertical line in (d) indicates the averaged $t_\mathrm{ot}$ of the corresponding ensemble of results.}
	\label{sample_d200}
\end{figure*}

 As shown in Sect.~\ref{sec4_1}, a clustered distribution of cavities results in a more rapid SDT process (i.e., a smaller value of $t_\mathrm{ot}$) than that resulting from a random distribution. The question thus arises as to whether these clusters of small cavities behave effectively the same as larger unclustered cavities during an SDT process. In an attempt to answer this question, additional simulations with randomly distributed larger cavities were carried out. Sample plots of randomly distributed large cavities of $d_\mathrm{c}=\SI{200}{\micro\metre}$ and $d_\mathrm{c}=\SI{300}{\micro\metre}$ are provided in Fig.~\ref{sample_d200}(b) and (c), respectively, in comparison to the sample plot of clustered small cavities ($d_\mathrm{c}=\SI{100}{\micro\metre}$) shown in Fig.~\ref{sample_d200}(a). The mean spacing of the large-cavity distributions is increased proportionally so that the overall porosity is maintained at the same value. As shown in Fig.~\ref{sample_d200}, the volume (i.e., area in this two-dimensional system) of each cavity of $d_\mathrm{c}=\SI{200}{\micro\metre}$ is approximately representative of the mean cluster size of small cavities. For an incident shock of $p_\mathrm{I}=\SI{7.0}{\giga\pascal}$, an ensemble of ten simulations were performed for the cases with randomly distributed large cavities.\\

A scatter chart comparing the $t_\mathrm{ot}$ resulting from the cases with randomly distributed large cavities (magenta diamonds for $d_\mathrm{c}=\SI{200}{\micro\metre}$ and  green squares for $d_\mathrm{c}=\SI{300}{\micro\metre}$) to those with randomly distributed (red triangles) and clustered (blue circles) small cavities of $d_\mathrm{c}=\SI{100}{\micro\metre}$ is plotted in Fig~\ref{sample_d200}(d). Each vertical line indicates the mean $t_\mathrm{ot}$ for the corresponding ensemble of results. The results of $t_\mathrm{ot}$ for the cases with a random distribution of cavities of three different sizes scatter over approximately the same range from $\SI{3.8}{\micro\second}$ to $\SI{4.3}{\micro\second}$. The mean $t_\mathrm{ot}$ for these three cases fall within a small range from $\SI{4.05}{\micro\second}$ to $\SI{4.12}{\micro\second}$. The mean $t_\mathrm{ot}$ resulting from the case with clustered small cavities ($\SI{3.72}{\micro\second}$) is significantly less than those from the cases with randomly distributed larger cavities. 

\subsection{\label{sec4_4}Detailed views of shock-cluster interactions}

Further detailed views of the interactions between an incident shock wave and clustered cavities are provided in this subsection. In order to more clearly identify the unique features arising from a shock-cluster interaction, two simple scenarios with an incident shock wave passing over a pair of vertically and horizontally aligned, clustered cavities are considered. The resulting dynamics from each of these scenarios is compared with that from the case with a pair of correspondingly aligned, unclustered cavities. Note that, to ensure a clearer visualisation of the detailed wave dynamics, the small-scaled simulations of an incident shock interacting with an individual cluster reported in this subsection were performed with a finer computational grid of $\mathrm{d}x=\SI{0.25}{\micro\metre}$, i.e., $400$ computational cells across the diameter of each cavity.\\

\begin{figure*}
\centerline{\includegraphics[width=0.8\textwidth]{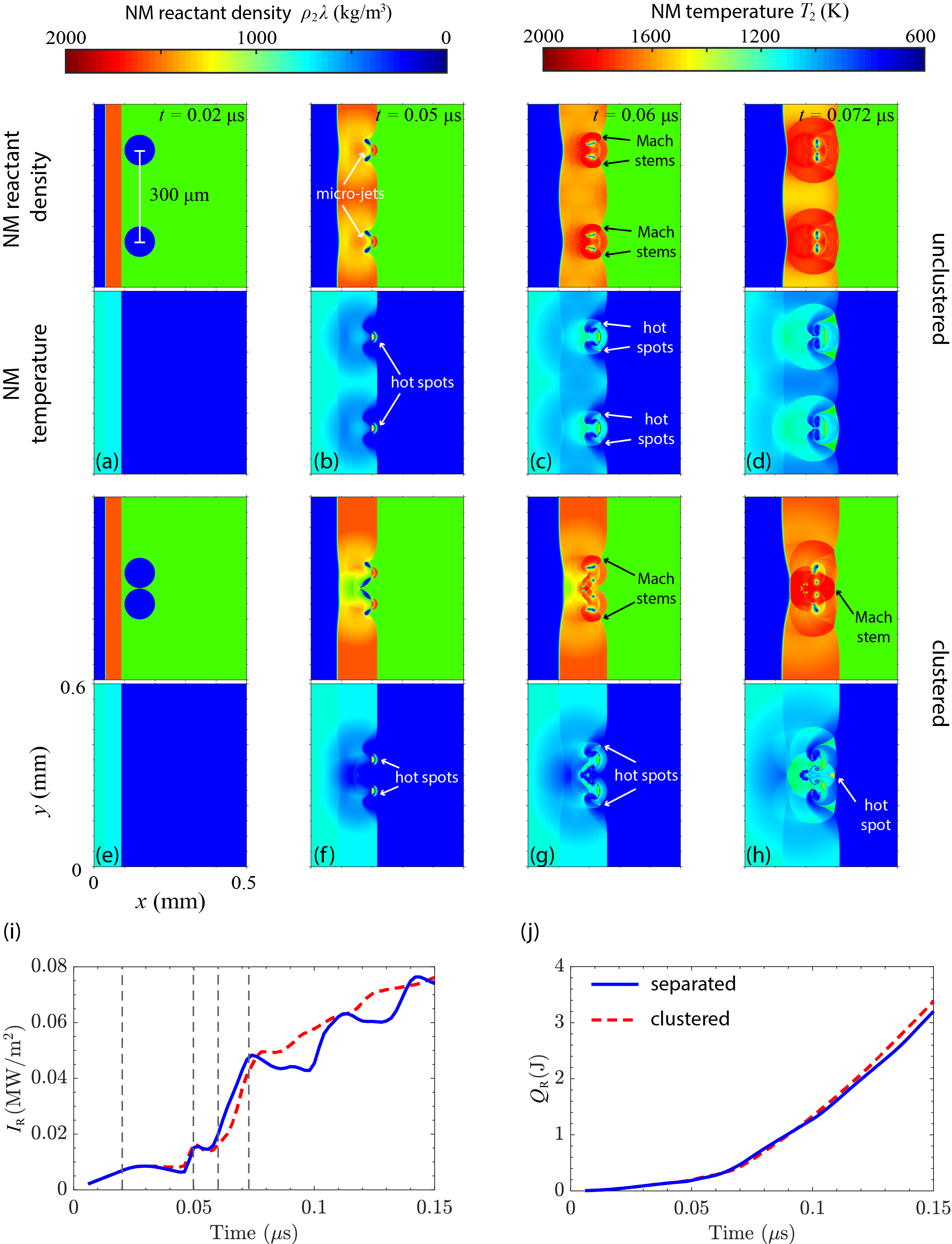}}
		\caption{A detailed view of the interaction between an incident shock wave and a pair of vertically aligned cavities, i.e., aligned perpendicular to the propagation direction of the incident shock. Subfigures (a)-(d) (in the upper row) are for the case with two cavities separated by a distance of $\delta_\mathrm{c}=\SI{300}{\micro\metre}$; subfigures (e)-(h) (in the lower row) are for the case with two clustered cavities. In each subfigure, the upper contour plot shows the field of NM reactant density $\rho_2 \lambda$, and the lower plot shows the NM temperature field $T_2$. Snapshots at four different times throughout the shock-cavity interaction are shown for each case. The specific rate of energy release per unit cross-sectional area of the explosive $I_\mathrm{R}$ and the amount of energy release $Q_\mathrm{R}$ resulting from each case are plotted as a function of time in (i) and (j), respectively. The vertical grey dashed lines in (i) indicate the times corresponding to the snapshots.}
	\label{doublet_V}
\end{figure*}

Figure~\ref{doublet_V} shows a detailed view of the interaction between an incident shock wave and a pair of vertically aligned cavities. The subfigures in the top row are for the case with two cavities separated by a distance of $\delta_\mathrm{c}=\SI{300}{\micro\metre}$; those in the bottom row are for the case with two clustered cavities. In each subfigure, the upper contour plot shows the field of NM reactant density $\rho_2 \lambda$, and the lower plot shows the NM temperature field $T_2$. Snapshots at four different times throughout the shock-cavity interaction are shown in Fig.~\ref{doublet_V} as four columns of subfigures. The specific rate of energy release per unit cross-sectional area of the explosive $I_\mathrm{R}$ and the amount of energy release $Q_\mathrm{R}$ resulting from each case is plotted as function of time in Fig.~\ref{doublet_V}(i) and (j), respectively. In both unclustered and clustered cases, upon the passage of the shock wave, a micro-jet is formed along the centreline and impacts the downstream (right) interface of each cavity. As indicated by the white arrows in Fig.~\ref{doublet_V}(b) and (f), hot spots are formed upon the impact of the shock-induced micro-jets. For an unclustered cavity, the curved, transmitted shock wave across the cavity interact the curved incident shock passing next to the cavity (i.e., above and below the cavity in the $y$-direction). This shock interaction forms two Mach stems, and each Mach stem induces a new hot spot away from the centreline of the cavity. As indicated in Fig.~\ref{doublet_V}(c), four Mach-stem-induced hot spots are thus formed in the case with two unclustered cavities. For a pair of vertically aligned, clustered cavities, three Mach-stem-induced hot spots are formed as shown in Fig.~\ref{doublet_V}(g) and (h). The hot spot in the middle is induced by the Mach stem formed due to the interaction between the two transmitted shock waves. As shown in Fig.~\ref{doublet_V}(i) and (j), the reaction fluxes $I_\mathrm{R}$ and energy release $Q_\mathrm{R}$ resulting from both of the vertically aligned scenarios remain within the same order of magnitude throughout the entire interaction with the shock wave; after the middle Mach-stem-induced hot spot emerges, the $I_\mathrm{R}$ for the case with clustered cavities (dashed red curve) slightly surpasses the reaction flux for the case with unclustered cavities (solid blue curve), leading to a slightly faster increase in $Q_\mathrm{R}$.\\

\begin{figure*}
\centerline{\includegraphics[width=0.8\textwidth]{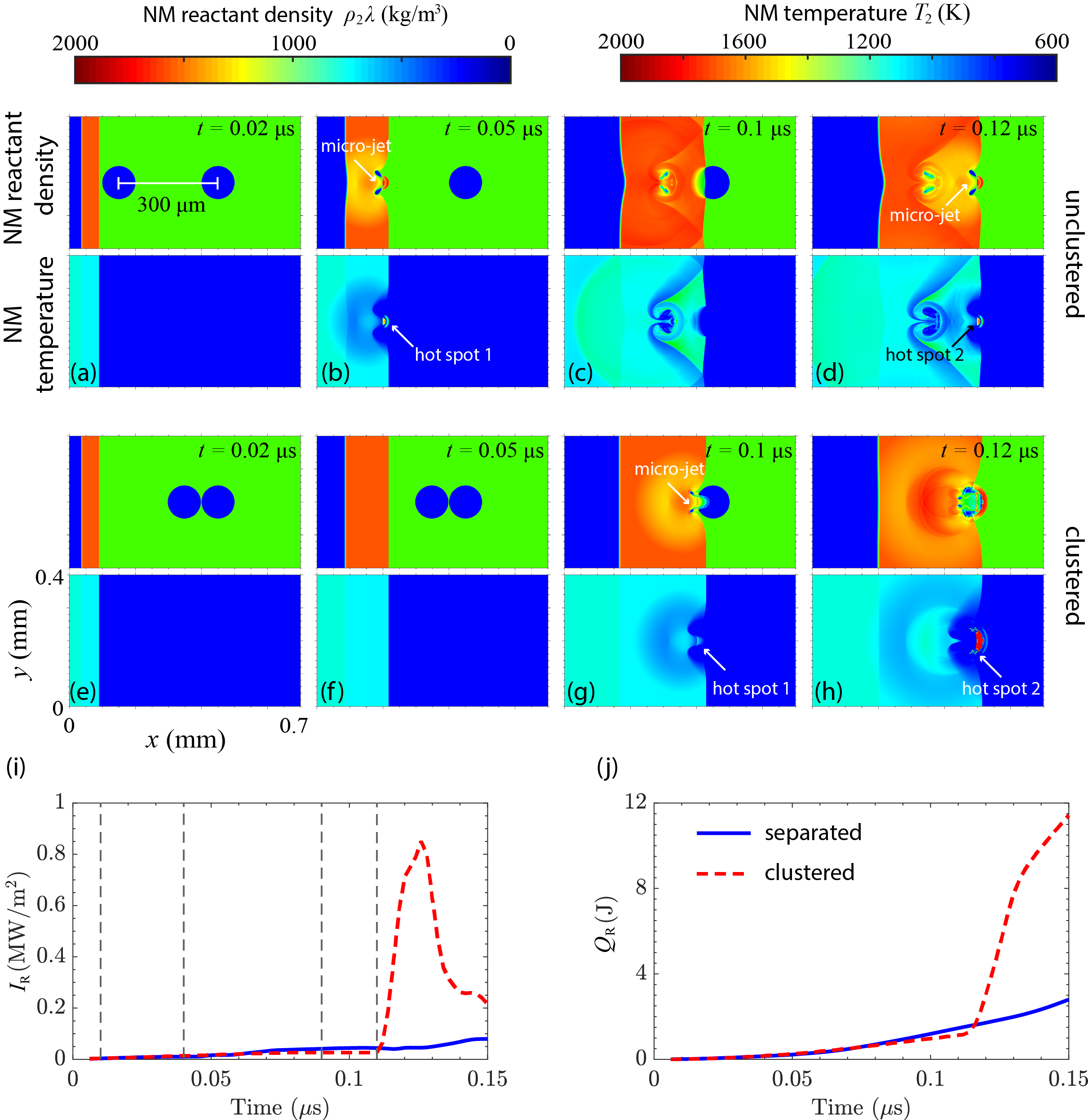}}
		\caption{A detailed view of the interaction between an incident shock wave and a pair of horizontally aligned cavities. Subfigures (a)-(d) (in the upper row) are for the case with two cavities separated by a distance of $\delta_\mathrm{c}=\SI{300}{\micro\metre}$; subfigures (e)-(h) (in the lower row) are for the case with two clustered cavities. In each subfigure, the upper contour plot shows the field of NM reactant density $\rho_2 \lambda$, and the lower plot shows the NM temperature field $T_2$. Snapshots at four different times throughout the shock-cavity interaction are shown for each case. The specific rate of energy release per unit cross-sectional area of the explosive $I_\mathrm{R}$ and the amount of energy release $Q_\mathrm{R}$ resulting from each case are plotted as a function of time in (i) and (j), respectively. The vertical grey dashed lines in (i) indicate the times corresponding to the snapshots.}
	\label{doublet_H}
\end{figure*}

Figure~\ref{doublet_H} shows a detailed view of the interaction between an incident shock wave and a pair of horizontally aligned cavities, i.e., aligned along the propagation direction of the incident shock. In the case with unclustered cavities separated by a distance of $\SI{300}{\micro\metre}$, the micro-jet-induced hot spot at each cavity occurs separately in time. The resulting reaction flux $I_\mathrm{R}$ (plotted as the solid blue curve in Fig.~\ref{doublet_H}(i)) remains of the same order of magnitude as those resulting from the cases with a pair of vertically aligned cavities shown in Fig.~\ref{doublet_V}(i). In the case with two horizontally aligned, clustered cavities, the micro-jet formed upon the collapse of the first (left) cavity penetrates into the second cavity (see Fig.~\ref{doublet_H}(g)); when this micro-jet strikes the downstream interface of the second cavity, a hot spot with a much more elevated temperature ($\approx \SI{2000}{\kelvin}$) is formed (see Fig.~\ref{doublet_H}(h)). As shown in Fig.~\ref{doublet_H}(i) and (j), the $I_\mathrm{R}$ and $Q_\mathrm{R}$ resulting from two horizontally clustered cavities increase by an order of magnitude after the second micro-jet-induced hot spot emerges.\\

\begin{figure*}
\centerline{\includegraphics[width=0.8\textwidth]{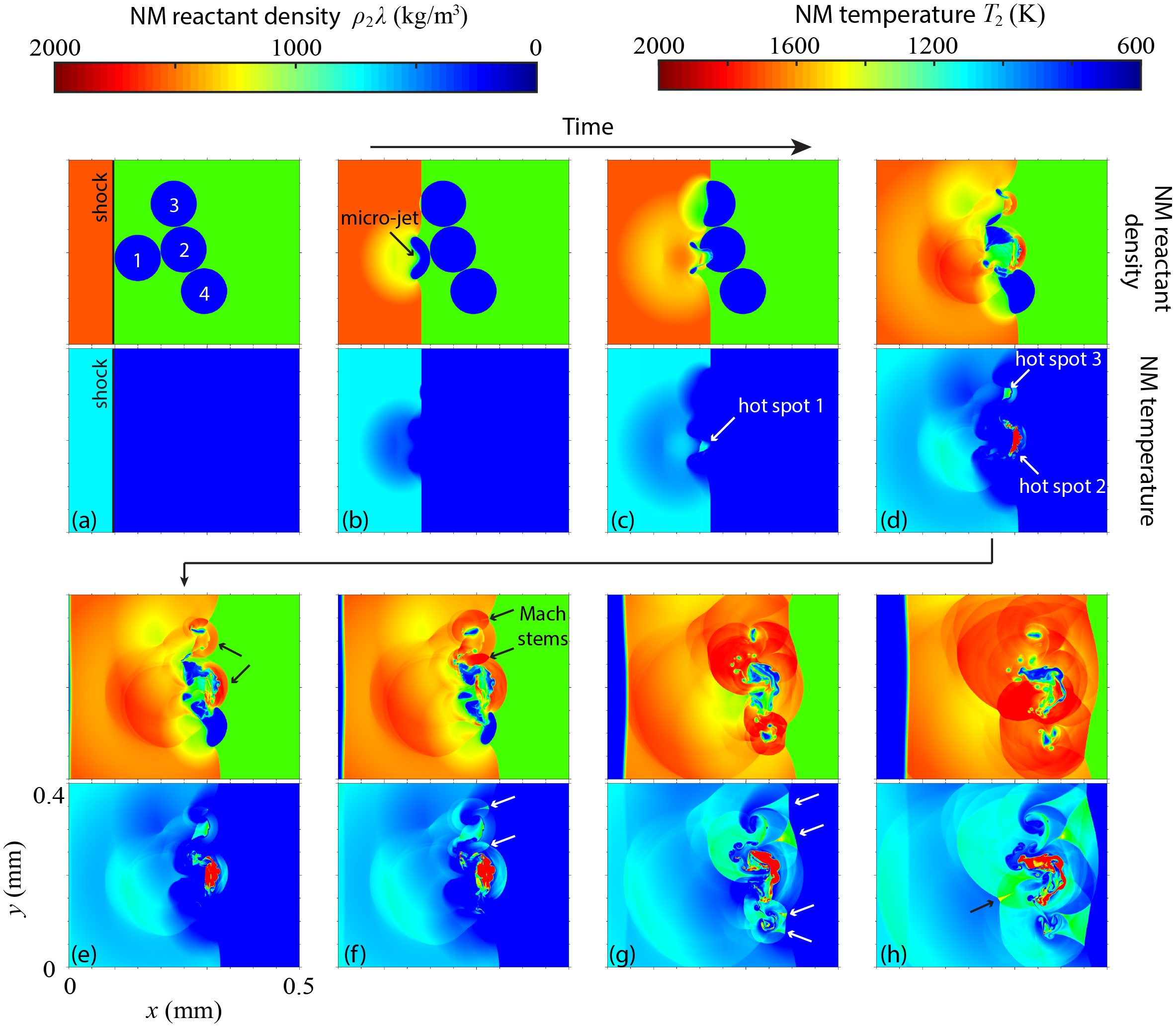}}
		\caption{A detailed view of the interaction between an incident shock wave and a typical cluster of four cavities. In each subfigure, the upper contour plot shows the field of NM reactant density $\rho_2 \lambda$, and the lower plot shows the NM temperature field $T_2$. Snapshots at eight different times throughout the shock-cluster interaction are shown in (a)-(h). Hot spots and other wave features are indicated by the arrows.}
	\label{cluster_zoom_in}
\end{figure*}

A detailed visualisation of the interaction between an incident shock wave and a sample cluster consisting of four cavities is provided in Fig.~\ref{cluster_zoom_in}. As the shock-induced micro-jets penetrate Cavities $1$ and $3$, Hot Spots $1$ and $3$ are formed as indicated in Fig.~\ref{cluster_zoom_in}(c) and (d), respectively. Given the fact that Cavities $1$ and $2$ are aligned along a direction nearly parallel to the propagation direction of the incident shock, the micro-jet penetrating through Cavity $2$ induces a hot spot with a much more elevated temperature, which is labelled as ``hot spot $2$'' in Fig.~\ref{cluster_zoom_in}(d). As indicated by the arrows in Fig.~\ref{cluster_zoom_in}(f) and (g), subsequent shock interactions further result in four Mach-stem-induced hot spots. Over the course of this sample shock-cluster interaction, seven hot spots can be identified within an area less than the square of the selected averaged spacing, i.e., ${\delta_\mathrm{c}}^2=\SI{300}{\micro\metre} \times \SI{300}{\micro\metre} = \SI{0.09}{\milli\metre\squared}$.\\

\section{\label{sec5}Discussion and comments}

\subsection{\label{sec5_1}Ranking in sensitising effect of various cavity distributions on the SDT process}

For the range of incident shock pressures (from $p_\mathrm{I}=\SI{7.0}{\giga\pascal}$ to $\SI{8.2}{\giga\pascal}$) considered in this paper, the detonation overtake time resulting from a mixture of NM and air-filled cavities is over $50\%$ less than that in neat NM (as shown in Fig.~\ref{Pop_plot}(a)), which is qualitatively in agreement with previous experimental findings~\cite{Dattelbaum2010}. This reduction in $t_\mathrm{ot}$ of an SDT process is known as the sensitising effect of the inclusion of mesoscale heterogeneities. In this range of shock pressure, the resulting SDT processes are thus classified as a hot-spot-driven regime. Via examining the results of $t_\mathrm{ot}$, the sensitising effect of various spatial distributions of cavities can be rank-ordered from the strongest to the weakest, i.e., from the shortest $t_\mathrm{ot}$ to the longest, as $\mathrm{Clustered} \rightarrow \mathrm{Random} \rightarrow \mathrm{Regular}$. The Pop-plot in Fig.~\ref{Pop_plot}(b) shows that this ranking in sensitising effect is consistently the same for the hot-spot-driven SDT processes over the considered range in $p_\mathrm{I}$.\\

The PDF of reaction rate as a function of $x$ at an early stage of the SDT (Fig.~\ref{pdf_x}) more directly reveals the effectiveness of hot spots created by shock interactions with heterogeneities. For all of the three distributions, the PDF of $\dot{q}$ spreads significantly above and below the bulk reaction rate in neat NM subjected to the same shock pressure (indicated by the horizontal dashed line in Fig.~\ref{pdf_x}). The distribution of $\dot{q}$ above the reaction rate in neat NM is due to the formation of hot spots. As the hot-spot-triggered release of energy proceeds, locally accumulated heat further increases the temperature, accelerating the reaction rate to $\dot{q} \geq 10^7 \, \mathrm{GW}/\mathrm{m^3}$, nearly two orders of magnitude greater than the bulk reaction rate. The ranking in sensitising effect discussed in the previous paragraph is related to how closely this fast-reaction zone (indicated by the red brackets in Fig.~\ref{pdf_x}) follows the leading shock front at an early time. At later times, the reaction rate is affected by the energy released from previously reacted material. The onward SDT progress is thus governed by a thermally positive feedback mechanism: As revealed by the time evolution of the PDF of reaction rate (Fig.~\ref{pdf_t}), a greater amount of hot-spot-induced fast-reacting material at an early time more rapidly increases the bulk reaction rate (associated with the maximum probability density), leading to a more rapid SDT process. 

\subsection{\label{sec5_2}Effect of randomness and clustering}

For a regular array of cavities, the averaged spacing between two neighbouring cavities $\delta_\mathrm{c}$ is equal to the mean distance from a cavity to its nearest neighbour $\overline{\delta}_\mathrm{min}$. In such a scenario, the hot spots created via shock-cavity interactions are regularly spaced, and the interactions among neighbouring hot spots are self-repetitive across the cross section of the explosive mixture. In the cases wherein the initial positions of cavities are randomised, the mean distance to the nearest neighbour is smaller than the averaged spacing of cavities, i.e., $\overline{\delta}_\mathrm{min} < \delta_\mathrm{c}$. Thus, there are hot spots that are more closely located than the regular spacing. For such a small time scale of an SDT process ($\sim \SI{1}{\micro\second}$), neighbouring hot spots interact most likely via pressure waves. The interaction among closely located hot spots further increases temperature and thus promotes the energy release. Subsequently, these hot spots merge and form high-temperature burnout kernels that are significantly larger than the spacing among cavities $\sim \delta_\mathrm{c}$. As shown in the zoom-in views of NM density and temperature fields at an early stage $t=\SI{2}{\micro\second}$ (Fig.~\ref{NM_density_temperature}), large burnout kernels are more populated in the case with clustered cavities than those in the case with a random distribution; only small burnout loci of sizes comparable to cavity diameter or averaged spacing appear in the case with a regular distribution. As shown in Fig.~\ref{NM_density}, reaction waves rapidly spread out from these high-temperature burnout kernels, consume the partially reacted explosive, and eventually catch up with the leading shock. These results suggest that the mean minimum distance among cavities, determining the probability of having closely located hot spots, is likely one of the limiting factors of the sensitising effect on an SDT process. As shown in Fig.~\ref{tot_rmin_pert}(c), $\overline{\delta}_\mathrm{min}$, a property describing the statistical nature of the initial distribution of cavities, characterises the sensitised SDT behaviour.\\

In the cases with clustered cavities, as revealed in detail in Sect.~\ref{sec4_4}, the hot spots are on-average closer to their neighbours than those resulting from a random distribution; shock-driven micro-jets penetrating multiple cavities aligned along the propagation direction of the incident shock wave give rise to hot spots with significantly greater temperatures. Clusters of more effective hot spots are thus formed, further enhancing the sensitising effect on an SDT process. As the results reported in Fig.~\ref{sample_d200}(d) suggest, these clusters of hot spots do not behave effectively the same as hot spots formed from unclustered, larger-sized cavities. The resulting ensemble-averaged $t_\mathrm{ot}$ does not significantly differ among the cases with randomly distributed cavities of three different sizes. The shock-induced collapse of larger-sized cavities produces larger hot spots. For a fixed overall porosity, however, the number density of these randomly distributed large hot spots is reduced due to a proportionally increased averaged spacing $\delta_\mathrm{c}$. Thus, the propensity for these hot spots to merge into large burnout kernels may not be significantly enhanced as the hot-spot size increases. This comparison suggests that clusters of finer heterogeneities (most likely formed in highly viscous liquid explosives) may have a more pronounced sensitising effect on the SDT process than that of unclustered, large-sized heterogeneities. It is of importance to note that the clusters considered in this study are relatively small, on-average consisting of three or four cavities. The sensitising effect may diminish as the mean cluster size increases to dozens or hundreds of cavities forming an energetically diluted porous core. Future effort is required to examine this speculated effect of large clusters of inert inclusions.\\ 

Although a two-dimensional system with simplified material EoS and idealized spatial distributions of circular cavities is considered in this study, the findings have some implications for experiments. As revealed in this work, a sensitised SDT process might be significantly influenced by the statistical nature of the mesoscale morphology of heterogeneities. In the experimental studies of gelled NM, the viscosity of the explosive matrix is affected by the properties and composition of different gelling agents. Depending on the viscosity of the gelled matrix and details in mixing procedure, the inert inclusions with similar properties (e.g., material, shape and size distribution, volume fraction) may cluster and exhibit statistically different distributions, giving rise to uncertainties in the macroscopic SDT parameters. For solid explosives, although the mechanisms of hot-spot formation and growth are more complex, a correlation between the statistical nature of the mesoscale structure and the sensitivity to shock initiation may exist. It is of importance to note that, as demonstrated by Michael and Nikiforakis~\cite{Michael2018_II}, the maximum hot-spot temperature resulting from a three-dimensional cavity collapse can be roughly $1.5$ times the peak temperature resulting from a two-dimensional scenario. This difference implies that the overall SDT times resulting from the current two-dimensional simulations might be significantly greater than the realistic value in three dimensions.\\

The main implication of the current findings is relevant to the development of meso-informed models of heterogeneous explosives. In meso-informed simulations, an explosive mixture with mesoscale heterogeneities is considered as a homogeneous medium with a reaction model that describes the rate of energy release as a function of local flow and thermodynamic properties. The conventional strategy to develop a meso-informed reaction model is based on calibrating the governing parameters against empirical data. Novel approaches using surrogate models that are machine-trained by ensembles of meso-resolved simulation data have recently been developed~\cite{Sen2018JAP,Nassar2019,Roy2019}. The current findings highlight the requirement of considering the statistical nature of heterogeneity distributions and clustering in order to develop high-fidelity meso-informed reaction models. This study thus offers a caveat that, considering only one or a small number of cavities, a mesoscale calculation might feed limited or biased information into a continuum-level simulation. Some recent modelling efforts have been made by Kittel~\textit{et al}.~\cite{Kittel2018PoF,Kittel2019PEP} and Bakarji and Tartakovsky~\cite{Bakarji2019APL} to link the statistical properties of mesoscale heterogeneities to continuum-level simulations of an explosive via a stochastic burn model. As suggested by the current study, the stochastic information fed into such a model can be obtained from meso-resolved simulations wherein a statistically significant distribution of heterogeneities is explicitly considered.\\

The use of simplistic models shall be explored to elucidate the more fundamental mechanisms linking the statistical nature of hot-spot distribution to the macroscopic SDT behaviours. Without modelling any detailed mechanisms of hot-spot formation and material EoS, Hill~\cite{Hill2002} developed a semi-analytic model (based on the statistical hot spot reactive model proposed by Nichols and Tarver~\cite{NicholsTarver2002}) in an attempt to capture the statistical nature of shock-triggered heterogeneous reactions. This model, wherein the growth rate of a hot spot is assumed to be independent of other hot spots, is conceptually more relevant to the later formed burnout kernels, but unable to capture the interacting hot spots formed upon the collapse of cavities. Via incorporating hot-spot interactions and statistically different types of distributions, this simplistic model might be used to further interpret the effect of mesoscale distributions revealed in the current study.

\section{\label{sec6}Conclusion}
Two-dimensional, meso-resolved simulations have been performed to examine the effect of statistically different spatial distributions of air-filled cavities on the shock-to-detonation transition in liquid nitromethane. It is found that, for a fixed overall porosity, a spatially random distribution of cavities (generated via a Poisson process) more effectively accelerates the SDT process than a regular array of cavities does. The sensitising effect on SDT can be further enhanced by the presence of small clusters of cavities. Via exploring a spectrum of scenarios ranging from a regular distribution to a random distribution, the mean distance of a cavity to its nearest neighbour $\overline{\delta}_\mathrm{min}$---a statistical property characterising the spatial randomness of a distribution of cavities---seemingly determines the sensitising effect of heterogeneities on an SDT process. For a distribution with a smaller $\overline{\delta}_\mathrm{min}$, closely located hot spots are more likely to occur, giving rise to a higher propensity to form large burnout kernels, and thus, resulting in a more rapid SDT process. Small clusters of several cavities, producing even more closely packed hot spots, further enhance the sensitising effect. For a fixed overall porosity, the sensitising effect of randomly distributed, unclustered cavities is not significantly changed as cavity size varies at least within an order of magnitude.

\begin{acknowledgments}
X.C.M. is supported by an NSERC Postdoctoral Fellowship (PDF-502505-2017). Computing resources used in this work were provided by Compute Canada and the University of Cambridge.
\end{acknowledgments}

\appendix*
\section{Simulation code validation}
The validation of the mathematical formulation used in this study for reacting, multiphase flows has been reported in detail by Michael and Nikiforakis~\cite{Michael2016JCP}. The current implementation of this model formulation invoking GPU-enabled parallel computing has been used in a previous study~\cite{Mi2019JAP}. However, the information with regards to the code validation was omitted in~\cite{Mi2019JAP}. A simulation of the shock-induced collapse of a single cavity in reactive liquid NM has been performed for validation purposes and presented in this Appendix. The results are compared to those reported in~\cite{Michael2018_I,Michael2018_II} wherein the same governing equations (including EoS and reaction rate model) were solved via an independently developed CPU-based simulation code.\\

\begin{figure*}
\centerline{\includegraphics[width=0.5\textwidth]{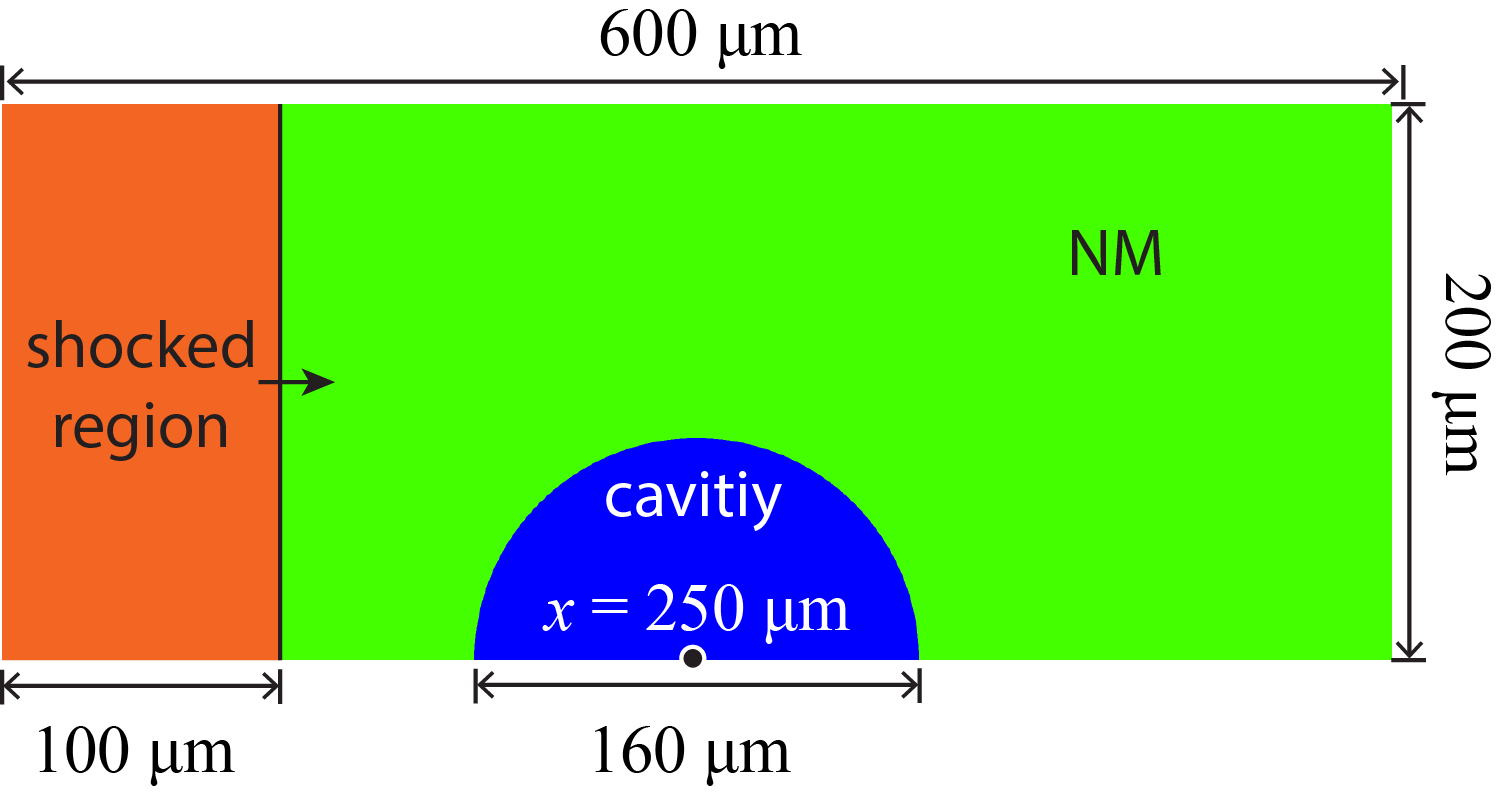}}
		\caption{A schematic showing the initial domain configuration of the validation problem: A single cavity with a diameter of $\SI{160}{\micro\meter}$ is subjected to an incident shock wave of $p_\mathrm{I}=\SI{10.98}{\giga\pascal}$.}
	\label{schematic_single}
\end{figure*}

\begin{figure*}
\centerline{\includegraphics[width=0.5\textwidth]{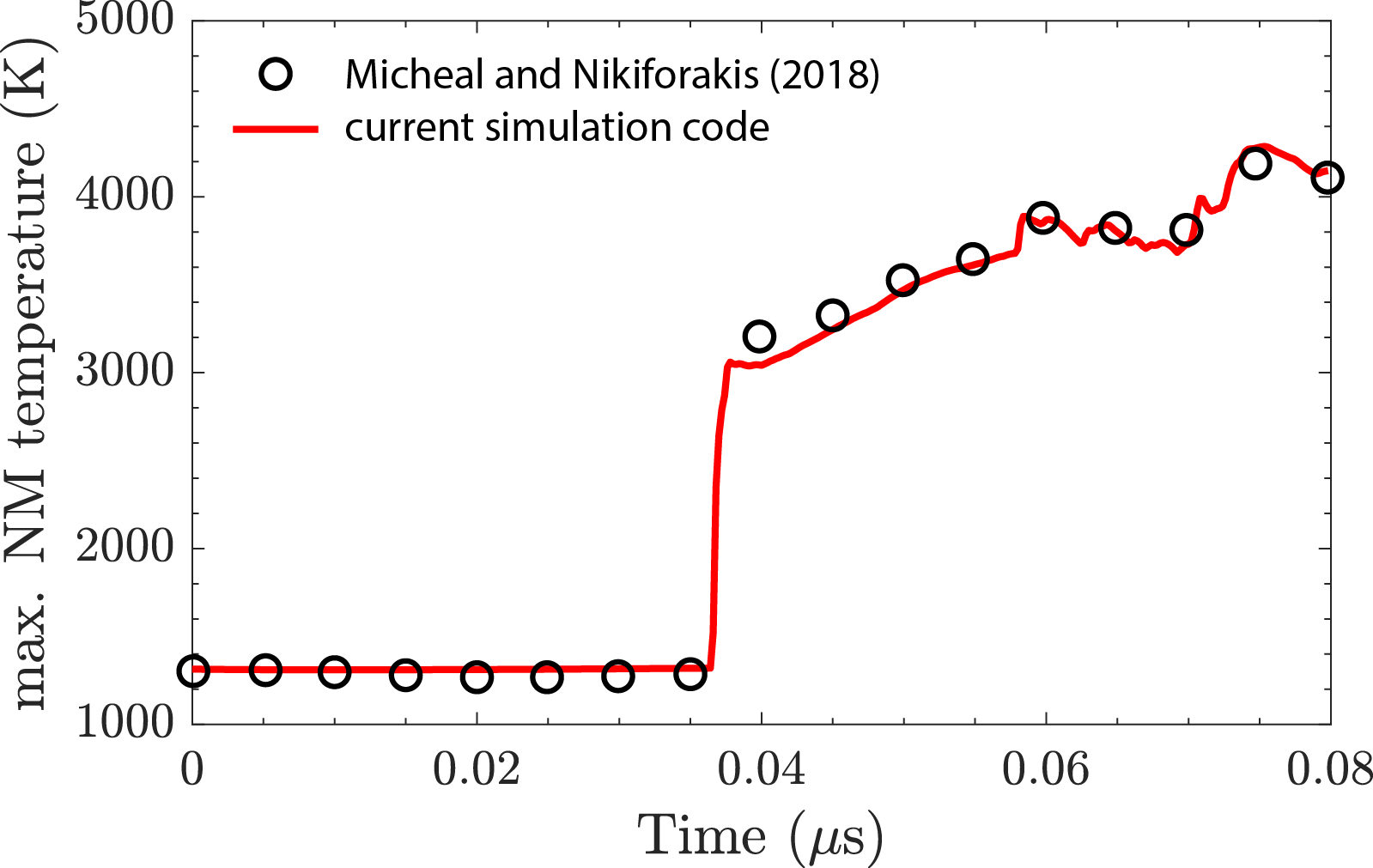}}
		\caption{Comparison between the current simulation result (red curve) of maximum hot-spot temperature in NM as function of time and that reported in literature~\cite{Michael2018_II} (black circles).}
	\label{Tmax_single}
\end{figure*}

Figure~\ref{schematic_single} shows the initial configuration of the computational domain for the single-cavity simulations. The dimensions of the initial features are the same as those reported in~\cite{Michael2018_I,Michael2018_II}. A cavity with a diameter of $\SI{160}{\micro\meter}$ is subjected to an incident shock wave of $p_\mathrm{I}=\SI{10.98}{\giga\pascal}$. The simulation was performed at a grid resolution of $\mathrm{d}x=\mathrm{d}x=\SI{0.3125}{\micro\meter}$, consistent with that used in~\cite{Michael2018_II}.  In Fig.~\ref{Tmax_single}, the resulting maximum temperature of NM in the domain, i.e., maximum hot-spot temperature, as a function of time is compared to Michael and Nikiforakis' simulation result reported in Fig.~14 of~\cite{Michael2018_II}. A good agreement between the current simulation result and that in the literature is demonstrated in Fig.~\ref{Tmax_single}.

\bibliography{detonation}

\end{document}